\documentclass[%
 reprint,
superscriptaddress,
amsmath,amssymb,
aps,
prx,
]{revtex4-2}

\usepackage{graphicx}
\usepackage{dcolumn}
\usepackage{bm}
\usepackage{hyperref}
\usepackage{booktabs}
\usepackage{xcolor}

\usepackage{xcolor,cancel}

\begin{document}
\preprint{APS/123-QED}

\title{An AI-powered Technology Stack for Solving Many-Electron Field Theory}
\author{Pengcheng Hou}
\email{houpc@hfnl.cn}
\affiliation{Hefei National Laboratory, University of Science and Technology of China, Hefei 230088, China}

\author{Tao Wang}
\email{taowang@umass.edu}
\affiliation{Department of Physics, University of Massachusetts, Amherst, MA 01003, USA}

\author{Daniel Cerkoney}
\email{dcerkoney@physics.rutgers.edu}
\affiliation{Department of Physics and Astronomy, Rutgers, The State University of New Jersey, Piscataway, NJ 08854-8019 USA}

\author{Xiansheng Cai}
\affiliation{Department of Physics, University of Massachusetts, Amherst, MA 01003, USA}
\author{Zhiyi Li}
\affiliation{Department of Modern Physics, University of Science and Technology of China, Hefei, Anhui 230026, China}
\author{Youjin Deng}
\email{yjdeng@ustc.edu.cn}
\affiliation{Department of Modern Physics, University of Science and Technology of China, Hefei, Anhui 230026, China}
\affiliation{Hefei National Laboratory, University of Science and Technology of China, Hefei 230088, China}
\author{Lei Wang}
\email{wanglei@iphy.ac.cn}
\affiliation{Beijing National Laboratory for Condensed Matter Physics and Institute of Physics, \\Chinese Academy of Sciences, Beijing 100190, China}
\author{Kun Chen}
\email{chenkun@itp.ac.cn}
\affiliation{CAS Key Laboratory of Theoretical Physics, Institute of Theoretical Physics, Chinese Academy of Sciences, Beijing 100190, China}
\affiliation{Department of Physics and Astronomy, Rutgers, The State University of New Jersey, Piscataway, NJ 08854-8019 USA}
\affiliation{Center for Computational Quantum Physics, Flatiron Institute, 162 5th Avenue, New York, New York 10010}

\date{\today}

\begin{abstract}
    Quantum field theory (QFT) for interacting many-electron systems is fundamental to condensed matter physics, yet achieving accurate solutions confronts computational challenges in managing the combinatorial complexity of Feynman diagrams, implementing systematic renormalization, and evaluating high-dimensional integrals. We present a unifying framework that integrates QFT computational workflows with an AI-powered technology stack.
    A cornerstone of this framework is representing Feynman diagrams as computational graphs, which structures the inherent mathematical complexity and facilitates the application of optimized algorithms developed for machine learning and high-performance computing.
    Consequently, automatic differentiation, native to these graph representations, delivers efficient, fully automated, high-order field-theoretic renormalization procedures. This graph-centric approach also enables sophisticated numerical integration; our neural-network-enhanced Monte Carlo method, accelerated via massively parallel GPU implementation, efficiently evaluates challenging high-dimensional diagrammatic integrals.
    Applying this framework to the uniform electron gas, we determine the quasiparticle effective mass to a precision significantly surpassing current state-of-the-art simulations.
    Our work demonstrates the transformative potential of integrating AI-driven computational advances with QFT, opening systematic pathways for solving complex quantum many-body problems across disciplines.
\end{abstract}

\maketitle

\tableofcontents

\section{Introduction}
\label{sec:introduction}

Rapid advances in artificial intelligence (AI) are fundamentally reshaping scientific research, offering systematic tools and new paradigms to address formerly intractable problems.
Although specific AI methodologies such as neural networks have garnered significant attention, the comprehensive technology stack for AI (AI tech stack)---encompassing differentiable programming and heterogeneous computing alongside neural networks---represents a development of potentially greater transformative power. This integrated ecosystem provides a versatile computational toolkit with applications extending well beyond the conventional ``AI for science" paradigm.

The capabilities of the complete AI tech stack invite a reevaluation of computational approaches in domains confronting fundamental limitations. Quantum field theory (QFT) applied to many-electron systems, a cornerstone of modern condensed matter physics and materials science, presents such a case. Despite its central importance for understanding electron correlations, transport phenomena, and exotic phases of matter, the quantitative application of perturbative QFT to these complex systems is severely constrained by computational barriers. Although initial applications of AI have shown promise, a systematic and comprehensive utilization of the entire AI tech stack holds the key to substantial, currently underappreciated, opportunities for progress.

Feynman diagrams constitute the primary language of perturbative QFT, offering both an intuitive visualization of particle interactions and a rigorous framework for calculating physical observables. However, achieving quantitative precision in many-electron systems via this route involves a sequence of computationally demanding stages, each presenting formidable challenges. First, the generation and enumeration of diagrams become difficult; as the perturbation order increases, the number of unique Feynman diagrams grows factorially, rapidly overwhelming any brute-force summation and limiting the accessible orders of theory~\cite{negele1998quantum}. Second, field-theoretic renormalization, essential for managing divergences and systematically resumming diagrammatic series to improve convergence, introduces further complexity, as traditional schemes often necessitate numerous counterterms, thereby amplifying the diagrammatic burden. Third, each diagrammatic contribution requires the evaluation of a high-dimensional integral over internal degrees of freedom, a task made numerically challenging by complex integrands, potential singularities, and high dimensionality. These interconnected challenges—diagram generation, renormalization, and integration—form a computational pipeline where bottlenecks at any stage compound dramatically, collectively 
limiting the predictive power of QFT in many-electron physics.

The modern AI tech stack offers a synergistic suite of capabilities well-suited to address these sequential challenges. Foundational to harnessing this synergy is the adoption of computational graphs, a core representational and execution paradigm underpinning modern AI frameworks (e.g., JAX~\cite{jax2018github}, PyTorch~\cite{NEURIPS2019_9015})
These directed acyclic graphs define the sequence of mathematical operations and data dependencies—from elementary interactions to complex diagrammatic sums. The significance of this representation lies in its enabling role within the AI tech stack: it provides the essential structure for systematic graph optimizations (such as common subexpression elimination~\cite{CSE1, CSE2} crucial for shared sub-diagrams), the precise application of automatic differentiation (AD) for high-order derivatives~\cite{baydin2018automatic,Taylor1}, and efficient, parallelized execution across heterogeneous hardware. While the inherent network structure of Feynman diagrams means that advanced QFT methodologies have long utilized graph-based reasoning implicitly or explicitly~\cite{diagMC4, PhysRevB.97.085117,PhysRevB.100.121102,Rossi_2020,kchen,wdm, taheridehkordi_algorithmic_2019,burke_torchami_2023,kozik_combinatorial_2023,sturt2024exploitingparallelismfastfeynman}, formalizing these expansions as computational graphs allows us to directly apply these powerful, well-developed capabilities from the AI domain. Building upon this structured foundation, differentiable programming (via AD on these graphs) facilitates automated field-theoretic renormalization. Subsequently, neural networks, particularly architectures such as normalizing flows (NFs)~\cite{NF2015,kobyzev_normalizing_2021}, offer a promising avenue to effectively tackle high-dimensional integration hurdles, using adaptive importance sampling, as demonstrated in related scientific fields~\cite{muller_neural_2019,gao_i-_2020,PhysRevD.101.076002,normalization_flow}. However, these prior explorations of AI-related techniques in QFT have largely focused on applying individual components to specific stages of the workflow; a comprehensive, integrated approach using the AI tech stack to address the entire pipeline remains inadequately explored.

\begin{figure}
    \centering
    \includegraphics[width=0.9\linewidth]{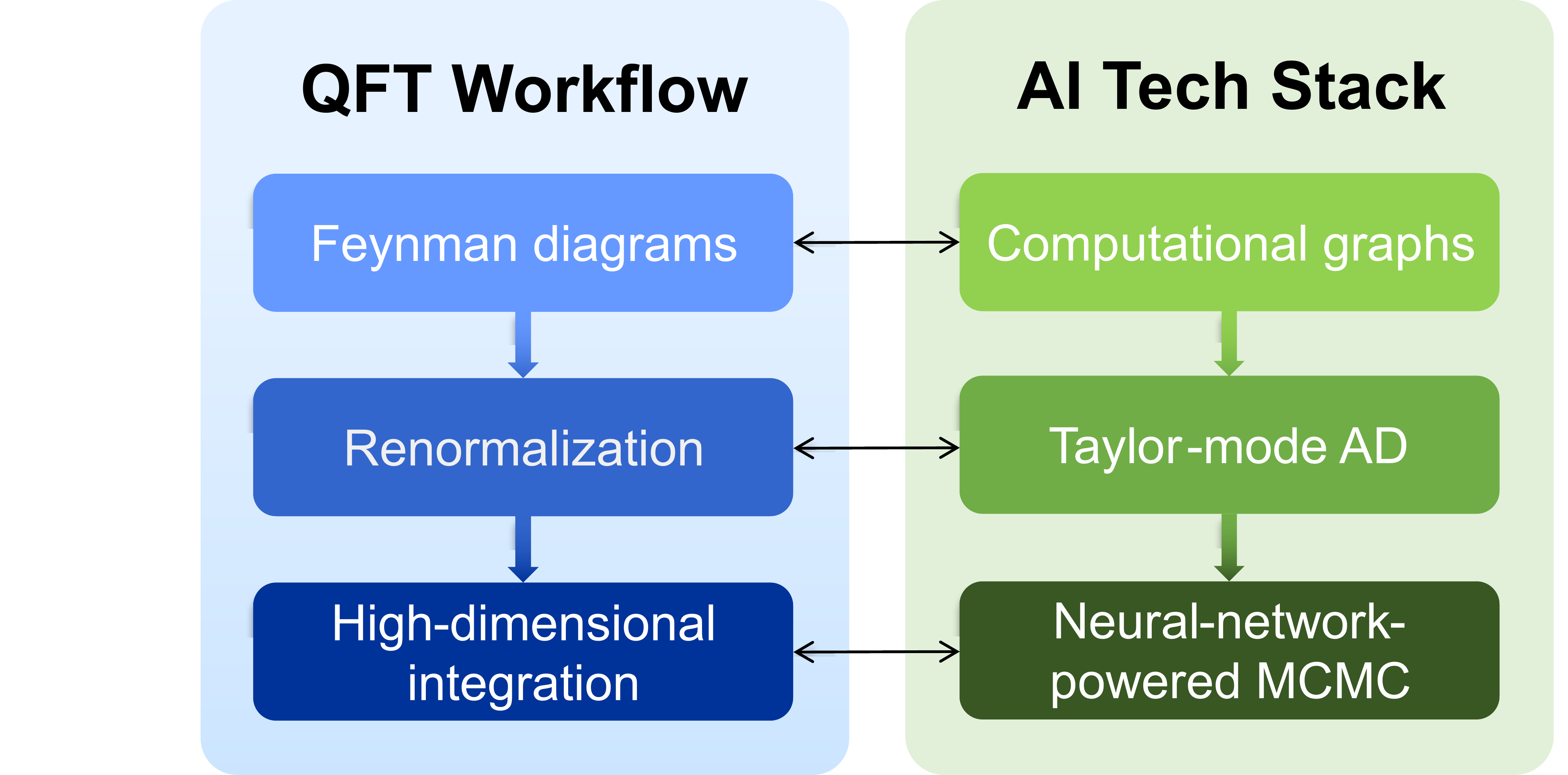}
    \caption{Mapping between the quantum field theory (QFT) workflow for many-electron systems and artificial intelligence (AI) computational methods. Left: The QFT calculation pipeline progresses from Feynman diagram construction, through field-theoretic renormalization, to high-dimensional integration of diagrammatic contributions. Right: Each stage in the QFT workflow is systematically addressed by a corresponding component of the AI tech stack: (1) computational graphs for efficient representation and management of diagrammatic series, (2) Taylor-mode automatic differentiation (AD) for precise and automated renormalization, and (3) Neural-network-powered Markov Chain Monte Carlo (MCMC) methods, particularly using normalizing flows, for accurate high-dimensional integration. This mapping forms the basis of our integrated framework for advancing many-electron QFT calculations.}
    \label{fig:mapping}
\end{figure}

In this work, we introduce an integrated framework that employs synergistic AI capabilities (as previously outlined) to systematically address computational challenges in QFT for many-electron systems, as shown in Fig.~\ref{fig:mapping}. Our methodology begins by constructing computationally efficient diagrammatic representations via compact computational graphs, derived from the perturbative interpretation of the self-consistent Dyson-Schwinger and parquet equations. These graphs, applicable across diverse physical domains (e.g., space-time, momentum-frequency), are algorithmically generated to represent key two-, three-, and four-point vertex functions. The explicit encoding of shared subdiagrams within this inherently hierarchical structure avoids redundant computation, substantially reducing the effective complexity of high-order series, thus formalizing and extending previous graph-based approaches through the systematic construction and optimization tools of the AI tech stack.

Building upon this optimized graph representation, we implement automated field-theoretic renormalization using Taylor-mode automatic differentiation~\cite{Taylor1, Taylor2, Taylor3}. This particular AD technique is well-suited for systematically computing the high-order functional derivatives required for renormalization directly on the computational graphs. We demonstrate its capacity to automate the precise calculation and incorporation of counterterms, thereby supporting an improved computational cost scaling of renormalized diagrams (e.g., from exponential to sub-exponential with respect to the differential order of these counterterms).

Finally, to address the computationally intensive integration stage, we develop and employ high-dimensional integration strategies incorporating normalizing flows within Markov chain Monte Carlo (MCMC) methods. This approach capitalizes on the ability of NFs to learn highly adaptive proposal distributions, which is essential for efficiently sampling the complex, often sharply peaked integrands characteristic of Feynman diagrams, thereby tackling a primary bottleneck in achieving high numerical precision.

Our unified implementation combines these components within a compiler-based framework designed for heterogeneous computing architectures (CPUs and GPUs). This compiler translates abstract Feynman diagrammatic structures into optimized computational graphs, directly incorporates Taylor-mode AD for renormalization procedures, and generates executable code compatible with high-performance machine learning (ML) libraries such as JAX and PyTorch. Such an architecture facilitates efficient parallel processing of both diagram evaluation and MCMC sampling, characterized by minimal control-flow branching.

We demonstrate the effectiveness of our integrated framework through high-precision calculations of the uniform electron gas (UEG) effective mass ratio. This application serves as a testbed and showcases how our approach---wherein efficient graph representations provide the computational backbone, Taylor-mode AD enables systematic and precise renormalization, and our developed integration methods ensure reliable numerical results---facilitates perturbative calculations of a high standard of accuracy. Specifically, we achieve a level of precision that surpasses outcomes from other leading contemporary many-body calculations by nearly two orders of magnitude~\cite{azadi2023,holzmann2023static}.

The methodology presented herein introduces a systematic integration of the AI tech stack into the fabric of many-electron field theory calculations. We believe that this work opens promising avenues for leveraging AI techniques to achieve further advances in quantum field theory and many-body physics research.

The paper is organized as follows: Section II details the computational graph representation of Feynman diagrams and the corresponding construction algorithms. Section III presents our implementation of field-theoretic renormalization using Taylor-mode AD. Section IV introduces our MCMC framework enhanced by NFs. Section V details the computational framework implementation. Section VI demonstrates our methodology through effective mass calculations for the UEG model. We conclude in Section VII by summarizing our work and discussing future developments and potential applications beyond the immediate scope of this work.

\section{Feynman Diagrams as Computational Graphs}
\label{sec:computgraph}
\subsection{Feynman diagrams}
\label{subsec:perturbation}

Feynman diagrams are fundamental to quantum many-body physics, providing both a graphical representation and a computational framework for QFTs. These diagrams are particularly vital in scenarios where exact solutions are elusive and where interaction terms are small relative to kinetic terms. They provide a visual representation of the perturbative expansion of the action in powers of the interaction, linking fundamental interactions with physical observables in a clear and insightful manner.

In QFT, the action $S$ encapsulates the dynamics and interactions of particle fields. It can be formulated in various domains, such as momentum and imaginary time, or in alternative combinations like space-time, space-frequency, momentum-time, and momentum-frequency. This is represented as:
\begin{equation}
    \label{eq:bare}
    S = \int_{\mathbf{k}\tau} \bar{\psi}_{\mathbf{k}\tau}\hat{g}^{-1}_{\mathbf{k}\tau}\psi_{\mathbf{k}\tau} + \int_{\mathbf{k}\mathbf{k'}\mathbf{q}\tau} V_\mathbf{q} \bar{\psi}_{\mathbf{k}+\mathbf{q}\tau}\bar{\psi}_{\mathbf{k'}-\mathbf{q} \tau}\psi_{\mathbf{k'} \tau}\psi_{\mathbf{k} \tau}.
\end{equation}
Here, the integration measures imply integration over internal momenta $\mathbf{k}$, $\mathbf{k^\prime}$, and $\mathbf{q}$ and imaginary time $\tau$, while $\bar{\psi}$ and $\psi$ are either bosonic or fermionic Grassmann fields. The bare propagator of the particle is given by
\begin{equation}
    \label{eq:propagator}
    \hat{g}_{\mathbf{k}\tau}^{-1} = \frac{\partial}{\partial \tau} +\epsilon_\mathbf{k} \quad \leftrightarrow \quad \hat{g}_{\mathbf{k}\omega_n}^{-1} = -i\omega_n +\epsilon_\mathbf{k},
\end{equation}
where $\epsilon_\mathbf{k}$ is the single-particle energy dispersion and $\omega_n$ is the Matsubara frequency. For simplicity, we consider an interaction potential $V_\mathbf{q}$ dependent only on momentum transfer $\mathbf{q}$ (e.g., Coulomb interaction). However, the methodologies developed in this paper are adaptable to more complex interaction forms, including those with frequency dependence or non-locality.

Physical quantities in QFT are calculated perturbatively as an expansion in powers of the interaction strength $V_\mathbf{q}$. Feynman diagrams provide a systematic method to construct and visualize each term in this expansion. Each diagram depicts a specific sequence of events: particles propagate (represented by lines) and interact (represented by points where lines meet, called vertices). Well-defined `Feynman rules' translate these graphical elements into precise mathematical expressions. Specifically, lines correspond to particle propagators, $\hat{g}$ (defined in Eq.~\eqref{eq:propagator}), and vertices correspond to interaction terms, $V_\mathbf{q}$.

\begin{figure}
    \includegraphics[width=\columnwidth]{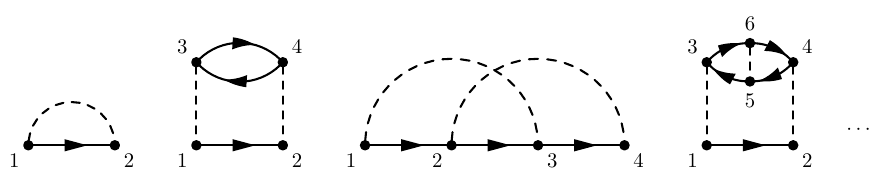}
    \caption{Feynman diagrams in the bare expansion of the self-energy. The bare propagator $g$ and interaction $V$ are represented by directed and dashed lines, respectively. We use an abbreviated notation $1 \equiv  \tau_1$ to denote the indices of the imaginary time variables. We omit the momentum variables for simplicity.}\label{fig:sigma_diags_bare}
\end{figure}

A key quantity derived from this framework is the self-energy, $\Sigma$, which describes how one-electron properties, such as energy and lifetime, are modified by interactions with the surrounding many-body system. Figure~\ref{fig:sigma_diags_bare} illustrates some low-order diagrams contributing to the self-energy. In $d$ spatial dimensions, the $n$th-order perturbation of the self-energy $\Sigma^{(n)}$ consists of Feynman diagrams with $n$ instantaneous interaction lines ($2n$ vertices), representing an $n (d + 1)$-dimensional integration,
\begin{equation}
    \Sigma^{(n)} = \int d\mathcal{V} \sum_{t\in \mathcal{T}_n}\, W_t^{(n)}(\mathcal{V}) \label{eq:sigma_integration}
\end{equation}
with measure $d\mathcal{V} = \prod^{n}_{i=1} d^{d}k_i d\tau_i$ and internal variables $\mathcal{V} = (\mathbf{k}_1, \cdots, \mathbf{k}_n; \tau_1, \cdots, \tau_n)$. Here the summation is over the set of all $n$th-order diagram topologies $\mathcal{T}_n$ (see Fig.~\ref{fig:sigma_diags_bare}) which grows as $\mathcal{O}(n!)$~\cite{diagMC4}.

This inherent combinatorial complexity and the high dimensionality of the integrations have historically constrained perturbative QFT applications in condensed matter to relatively low orders. Landmark examples include the GW approximation~\cite{hedin,godby1988self,hybertsen1986electron} for electronic quasiparticle energies and the Eliashberg theory of superconductivity~\cite{eliashberg1,eliashberg2}, both of which are fundamentally limited in the diagrammatic orders they incorporate. While providing crucial insights, such approximations are often insufficient for systems exhibiting strong correlations or where higher-order effects dictate the physical behavior.

The need to transcend these limitations has driven the development of sophisticated numerical methods. Diagrammatic Monte Carlo (DiagMC) techniques~\cite{diagMC1, diagMC2, Kozik_2010, diagMC4, boldDiagMC, kchen, rossi2018, van2012feynman, shifted_action} represent a significant advance, employing stochastic integration to sample the diagrammatic series and evaluate the high-dimensional integrals embodied in Eq.~\eqref{eq:sigma_integration}. DiagMC has enabled investigations of diverse correlated systems, such as the uniform electron gas~\cite{kchen,haule2022single,dynamic,wdm}, the unitary Fermi gas~\cite{van2012feynman,rossi2018,PhysRevLett.121.130406,PhysRevB.99.035140}, and Hubbard-type models~\cite{Kozik_2010,PhysRevLett.113.195301,deng_emergent_2015,PhysRevB.96.081117,PhysRevLett.124.117602,simkovic_extended_2020,PhysRevB.103.195147,iv_two-dimensional_2022,lenihan_evaluating_2022}. However, its efficacy can be hampered by the notorious fermion sign problem~\cite{troyer_computational_2005,Kozik_2010}, where statistical variance due to alternating signs limits achievable precision, particularly at low temperatures or strong coupling.

An alternative paradigm, Tensor Crossing Interpolation (TCI)~\cite{oseledets_tt-cross_2010,TCI1,TCI2}, utilizes tensor network representations, notably tensor train decompositions, to approximate the sum of Feynman diagrams. This method aims to recast the multivariate integration as a series of lower-dimensional operations, potentially mitigating issues associated with direct high-dimensional sampling. TCI has demonstrated efficacy in solving quantum impurity models~\cite{TCI3}.

Despite the conceptual differences between these advanced computational frameworks, a common and persistent bottleneck remains: the efficient construction and evaluation of the integrand $W_t^{(n)}$ for individual diagram topologies, repeated across a vast number of internal variable configurations $(\mathcal{V}_t)$ and for an exponentially growing set of topologies $\mathcal{T}_n$. Addressing this fundamental evaluation cost, alongside the combinatorial management of the diagrammatic series itself, is essential for pushing the frontiers of perturbative QFT.

\subsection{Computational graph representation}
\label{subsec:graphrep}

To address the combinatorial and evaluative complexities inherent in high-order perturbative QFT calculations, 
we adopt a structured representational framework inspired by advancements in managing large-scale computations in other domains. Specifically, we turn to computational graphs, a concept central to modern AI and ML platforms~\cite{tensorflow2015-whitepaper,NEURIPS2019_9015,jax2018github}. These graphs provide explicit encoding of mathematical operations and their data dependencies, enabling systematic analysis, optimization via graph transformations (e.g., common subexpression elimination, operator fusion), and efficient execution. The principles that render computational graphs effective in AI/ML—modularity, explicit data flow, and amenability to automated optimization—are directly applicable to the challenges of evaluating complex diagrammatic sums in QFT.
Our goal is to employ this framework to develop an efficient Feynman diagram representation that provides a tractable and optimizable foundation for perturbative QFT.

\begin{figure*}
    \centering
    \includegraphics[width=0.8\textwidth]{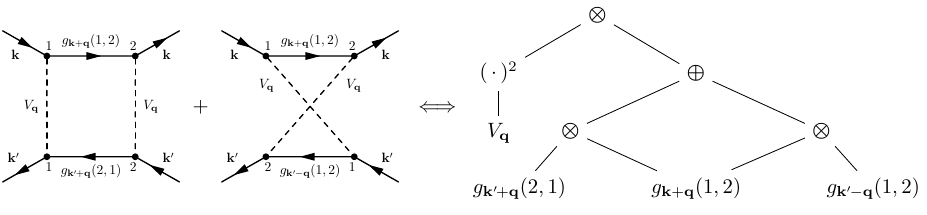}
    \caption{\label{fig:factorization}  Illustration of the computational graph of two 4-point vertex function Feynman diagrams. These diagrams are distinguished by the exchange of interaction lines. The graph integrates both momentum (edge-dependent) and imaginary-time (vertex-dependent) variables.
        Each node corresponds explicitly to either propagators/interactions (leaf nodes) or elementary mathematical operations such as multiplication ($\otimes$), addition ($\oplus$), and exponentiation ($(\,\cdot\,)^n$).
        Factorization of shared structures into a compact computational graph reduces redundancy in evaluating these diagrams.}
\end{figure*}

A computational graph is a directed acyclic graph where nodes represent mathematical operations or input variables, and edges signify the flow of data. The integrand sum central to Eq.~\eqref{eq:sigma_integration} can be directly mapped onto such a graph. Leaf nodes typically correspond to elemental QFT components like bare propagators $\hat{g}$ and interaction vertices $V$. Subsequent internal nodes execute the arithmetic operations (products, sums) dictated by the Feynman rules for each diagram topology, with the graph culminating in a root node representing the total integrand value. This explicit graph structure not only defines the calculation but also exposes it to algorithmic manipulation for optimization. Figure~\ref{fig:factorization} illustrates this, showing how distinct Feynman diagrams can be represented and, crucially, how shared sub-structures can be factorized into a more compact graph to reduce computational redundancy. This factorization is analogous to model optimization techniques employed in AI/ML compilers.

The factorization principle demonstrated in Fig.~\ref{fig:factorization}, whereby shared computational work is identified and unified, is fundamental to leveraging computational graphs for the full perturbative series sum. High-order diagrammatic expansions are inherently characterized by a high degree of such common topological sub-components, leading to extensive recurrence of identical mathematical expressions across myriad diagrams. The critical objective is thus to construct a globally optimized computational graph for the entire series, where all such redundancies are systematically eliminated. Achieving this level of comprehensive optimization, particularly for the complex and numerous diagrams encountered in momentum-frequency space formulations, however, introduces distinct algorithmic challenges beyond the general computational complexity of perturbation theory.

A primary impediment is the strategic assignment of internal loop variables. The algebraic form of a sub-diagram's contribution is critically dependent on the routing of its internal momenta and frequencies through the diagram's topology. Inconsistent variable assignments across physically equivalent sub-structures---those representing the same sequence of interactions and propagations---can obscure their underlying commonality by yielding algebraically dissimilar mathematical expressions. This, in turn, masks important opportunities for computational reuse and hinders effective factorization. Establishing an optimal internal-variable labeling scheme that maximizes the explicit recurrence of identical mathematical forms for these shared components across a vast ensemble of diagrams is consequently a formidable combinatorial task.

A further significant challenge arises in the subsequent identification and factorization of these common sub-diagrams, even when their recurrence is made explicit by appropriate variable assignment. This task is analogous to Common Subexpression Elimination (CSE) in compiler theory, wherein redundant computations are identified and their results reused~\cite{CSE1, CSE2}. However, standard CSE algorithms, though foundational, often lack the requisite scalability and specialization for the unique structural properties and sheer magnitude of computational graphs characterizing high-order QFT expansions.

It is noteworthy that significant progress in taming the complexity of diagrammatic sums has been achieved within the Diagrammatic Monte Carlo (DiagMC) framework through determinantal algorithms, such as the connected determinant DiagMC~\cite{diagMC4,PhysRevB.102.195122}.
These approaches often construct sums of diagrams as determinants of matrices built from single-particle propagators, thereby effectively reducing the computational scaling from factorial, $O(N!)$, to exponential, $O(e^N)$, in perturbation order $N$. Such determinantal methods have proven particularly powerful for problems formulated in a real space-time representation. However, many QFT problems, including those involving non-local interactions or those naturally analyzed via spectral properties, are more appropriately or necessarily treated in momentum-frequency space. For these ubiquitous scenarios, a different route to systematic optimization has been to identify sign-canceling groupings of diagrams based on fermionic sign-structure analysis~\cite{PhysRevB.103.115141}. This symmetry-based grouping provides a physical basis for the consistent assignment of internal loop variables across a cluster of diagrams to maximize topological sign cancellation. While this approach dictates which diagrams should be summed together, the efficient evaluation of the resulting complex group sum---by factorizing its many shared topological sub-components---presents its own computational challenge.

The methodology detailed in the subsequent sections directly addresses these graph-specific optimization challenges, particularly for momentum (or frequency) representations. We introduce a bottom-up construction algorithm that, by design, generates compact computational graphs by inherently avoiding redundant computations of shared sub-diagrams from their inception. This approach establishes a foundation for efficient evaluation of high-order perturbative series and represents a key component in a broader strategy to integrate effective elements from the AI tech stack into the QFT computational toolkit.

\subsection{Graph construction algorithms}
\label{subsec:algorithms}

\subsubsection{Perturbative Dyson-Schwinger equations}

\begin{figure}
    \includegraphics[width=0.9\columnwidth]{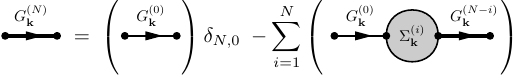}
    \includegraphics[width=0.6\columnwidth]{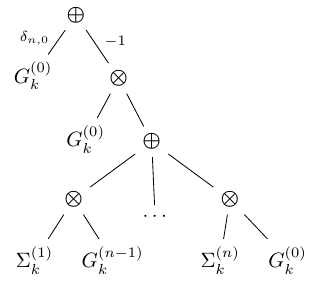}
    \caption{Construction of the computational graph for the $N$th-order Green's function $G_\mathbf{k}^{(N)}$. The top plot represents Eq.~\eqref{eq:Dyson}, illustrating the recursive definition of $G^{(N)}$ using self-energy contributions $\Sigma^{(i)}$ and lower-order Green's functions $G^{(N-i)}$. The schematic diagram below shows the computational graph of $G_\mathbf{k}^{(N)}$ based on the above equation, highlighting how intermediate nodes (marked by symbols $\otimes$ and $\oplus$ ) capture tensor operations reused throughout the computation, resulting in a compact representation.
    }
    \label{fig:green}
\end{figure}

The Dyson-Schwinger equations (DSEs) provide a fundamental non-perturbative framework in QFT for understanding particle interactions and propagation through exact relationships between Green's functions and vertex functions of varying particle numbers. In our approach, we apply the DSEs to generate compact computational graph representations for high-order Green's function and vertex function Feynman diagrams perturbatively. This approach exploits the hierarchical structure inherent in the DSEs to systematically construct and organize complex diagrammatic contributions, providing computational advantages for high-order perturbative expansions in QFT.

We demonstrate this approach by generating a computational graph for the $N$th-order Green's function from a set of $i$th-order self-energy diagrams ($i \leq N$) using the Dyson equation,
\begin{equation}
    G^{(N)}_\mathbf{k} = G^{(0)}_\mathbf{k}\delta_{N,0} - G^{(0)}_\mathbf{k}\sum^{N}_{i=1} \Sigma^{(i)}_\mathbf{k} G^{(N-i)}_\mathbf{k}.
    \label{eq:Dyson}
\end{equation}
For clarity, we present equations and figures with momentum labels only.

Notably, $N$-point vertex functions and Green's functions form multi-dimensional tensors when vertex-defined indices, such as imaginary-time and spin, are included. For instance, the Green's function in Fig.~\ref{fig:green}, with imaginary-time indices, becomes a matrix $G_\mathbf{q} \rightarrow \left[G_{\mathbf{q}}\right]_{\tau_i, \tau_j}$. In this tensorial representation, computational graph nodes represent tensors, with operations performed through tensor contractions rather than simple scalar multiplication. This formulation naturally expresses Feynman diagrams as a tensor network, providing a framework for systematic computational optimization.

\begin{figure}
    \includegraphics[width=0.85\columnwidth]{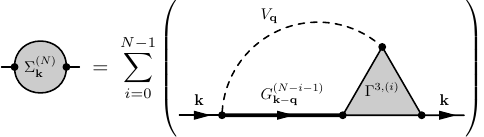}
    \includegraphics[width=0.7\columnwidth]{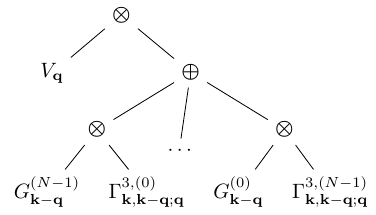}
    \caption{Construction of the computational graph for the $N$th-order self-energy $\Sigma_\mathbf{k}^{(N)}$. The top plot represents Eq.~\eqref{eq:sigma_pq}, illustrating the definition of $\Sigma^{(N)}$ using the full 3-point vertex functions $\Gamma^{3,(i)}$, Green's functions $G^{(N-i-1)}$, and the bare interaction $V$. The schematic diagram below shows the compact computational graph representation of $\Sigma_\mathbf{k}^{(N)}$ based on the above equation.    
    }
    \label{fig:self-energy}
\end{figure}

The self-energy contributions $\Sigma^{(i)}$, essential for constructing the Green's function $G^{(N)}$ via Eq.~\eqref{eq:Dyson}, are themselves recursively defined from lower-order quantities. Specifically, the $N$th-order self-energy $\Sigma^{(N)}$ is determined by the full 
3-point vertex function and Green's functions, as shown in Fig.~\ref{fig:self-energy} and given by the relation:
\begin{equation}
    \Sigma_{\mathbf{k}}^{(N)} = V_{\mathbf{q}} \sum^{N-1}_{i=0} G_{\mathbf{k}-\mathbf{q}}^{(N-i-1)}\Gamma^{3,(i)}_{\mathbf{k},\mathbf{k}-\mathbf{q};\mathbf{q}}, \label{eq:sigma_pq}
\end{equation}
where the integration over internal momenta is implied.

\begin{figure}
    \includegraphics[width=\columnwidth]{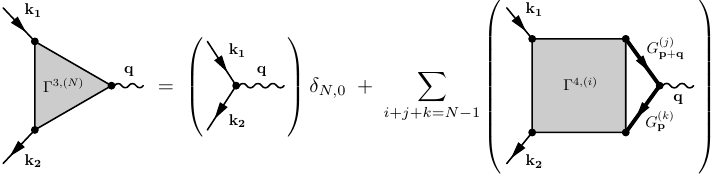}
    \includegraphics[width=\columnwidth]{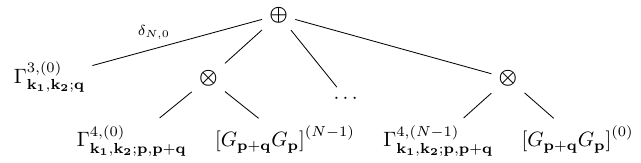}
    \includegraphics[width=0.5\columnwidth]{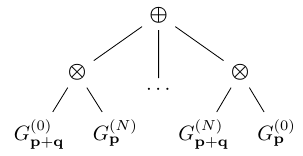}
    \caption{Construction of the computational graph for the $N$th-order 3-point vertex $\Gamma^{3,(N)}_{\mathbf{k}_1,\mathbf{k}_2;\mathbf{q}}$. The top plot represents Eq.~\eqref{eq:vertex3}, illustrating the definition of $\Gamma^{3,(N)}$ using the full 4-point vertex functions $\Gamma^{3,(i)}$ and the product of two Green's functions $\left[G_{\mathbf{p}+\mathbf{q}}G_\mathbf{p}\right]^{(N)}$. The schematic diagrams below show the compact computational graph representations of $\Gamma^{3,(N)}_{\mathbf{k}_1,\mathbf{k}_2;\mathbf{q}}$ and $\left[G_{\mathbf{p}+\mathbf{q}}G_\mathbf{p}\right]^{(N)}$, respectively.
    \label{fig:gamma3}
    }
\end{figure}

Following this hierarchical construction, the full 3-point vertex functions $\Gamma^{3,(i)}$, required for the self-energy in Eq.\eqref{eq:sigma_pq}, are in turn generated from the 4-point vertex. The $N$th-order 3-point vertex $\Gamma^{3,(N)}_{\mathbf{k}_1,\mathbf{k}_2;\mathbf{q}}$ is given by:
\begin{equation}
    \Gamma^{3,(N)}_{\mathbf{k}_1,\mathbf{k}_2;\mathbf{q}} = \Gamma^{3,(0)}_{\mathbf{k}_1,\mathbf{k}_2;\mathbf{q}}\delta_{N,0} + \sum^{N-1}_{i=0}\Gamma^{4,(i)}_{\mathbf{k}_1,\mathbf{k}_2;\mathbf{p}, \mathbf{p}+\mathbf{q}} \left[G_{\mathbf{p}+\mathbf{q}}G_{\mathbf{p}}\right]^{(N-i-1)},
    \label{eq:vertex3}
\end{equation}
where we define $\mathbf{q} \equiv \mathbf{k}_1 - \mathbf{k}_2$ for brevity, $\Gamma^{3,(0)}$ is the bare three-point vertex, and
\begin{equation}
    \left[G_{\mathbf{p}+\mathbf{q}}G_{\mathbf{p}}\right]^{(N)} = \sum^{N}_{j=0} G^{(j)}_{\mathbf{p}+\mathbf{q}} G^{(N-j)}_{\mathbf{p}}
\end{equation}
is the product of two Green's functions truncated to total order $N$. The generation of the computational graph for $\Gamma^{3,(N)}$ and its components is shown in Fig.~\ref{fig:gamma3}.
The full 4-point vertex function $\Gamma^{4,(i)}$, which provides the essential input to this perturbative DSE hierarchy, is constructed using the parquet equations as described in the following subsection.

\subsubsection{\label{sec:parquet} Perturbative parquet equations}

The parquet equations~\cite{parquet1, parquet2} are fundamental to describing interactions in complex systems through the comprehensive analysis and construction of 4-point vertex functions. These equations establish relationships between full vertex functions and their reducible and irreducible components across different interaction channels.

\begin{figure}
    \includegraphics[width=0.9\columnwidth]{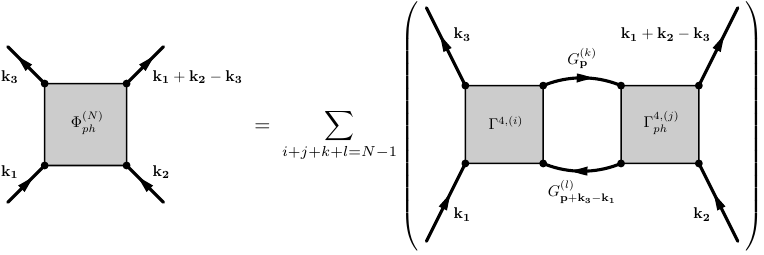}\\[2ex]
    \includegraphics[width=0.9\columnwidth]{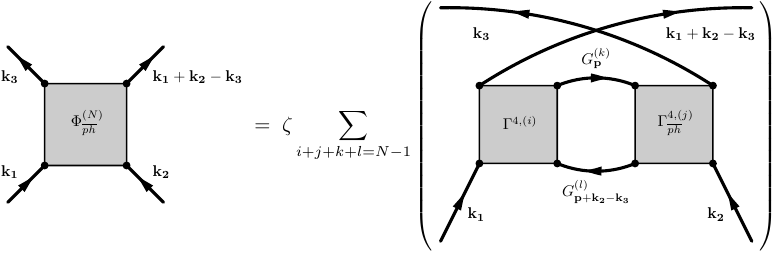}\\[2ex]
    \includegraphics[width=0.9\columnwidth]{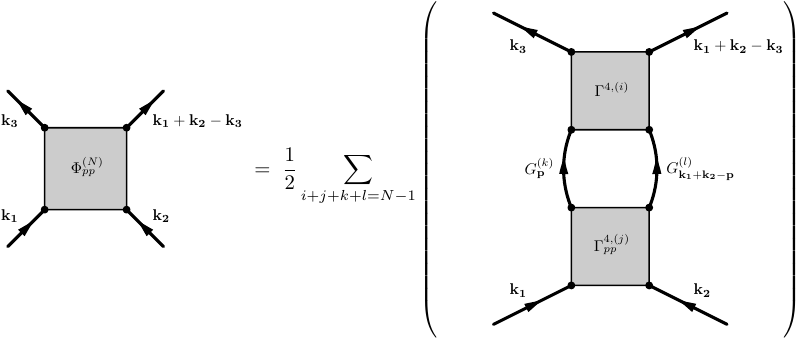}
    \caption{Diagrammatic construction of the $N$th-order reducible 4-point vertices $\Phi^{(N)}_{c}$ in each channel $c \in \{ph, \overline{ph}, pp\}$ following Eqs.~\eqref{eq:phi_ph}--\eqref{eq:phi_pp}. Here $\zeta = -1$ for fermions, and $+1$ for bosons.}\label{fig:phi_diags}
\end{figure}

\begin{figure}
    \centering
    \includegraphics[width=0.8\columnwidth]{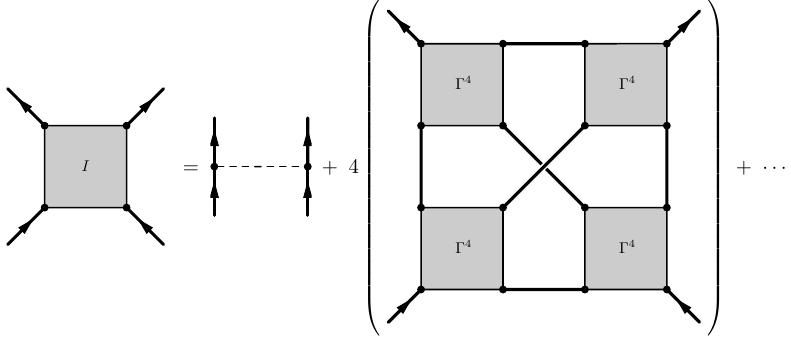}
    \caption{The fully-irreducible 4-point vertex $I$. The $N$th-order contribution is built bottom-up from the bare interaction $V$, Green's function $G$ and reducible 4-point vertex $\Gamma^4$ in an analogous manner to Eqs.~\eqref{eq:phi_ph}--\eqref{eq:phi_pp}. This figure illustrates one specific permutation of external legs; however, the complete representation of $I$ includes all possible permutations, not depicted here for clarity.}
    \label{fig:envelope_diag}
\end{figure}

While parquet equations traditionally yield self-consistent solutions for vertex functions~\cite{Smith1988,fRG2018}, we apply them perturbatively to construct the computational graph for the 4-point vertex Feynman diagrams. In this approach, each term in the parquet equations corresponds to a specific diagrammatic structure. The 4-point vertex function is constructed through the parquet equations,
\begin{equation}
    \Gamma^{4,(N)}_{\mathbf{k}_1, \mathbf{k}_2; \mathbf{k}_3, \mathbf{k}_4} = I^{(N)}_{\mathbf{k}_1, \mathbf{k}_2; \mathbf{k}_3, \mathbf{k}_4} + \sum_{c \in \{ph, \overline{ph}, pp\}} \Phi^{c,(N)}_{\mathbf{k}_1, \mathbf{k}_2; \mathbf{k}_3, \mathbf{k}_4},\label{eq:parquet_eqns}
\end{equation}
where $\mathbf{k}_4 \equiv \mathbf{k}_1 + \mathbf{k}_2 - \mathbf{k}_3$, $I^{(N)}$ denotes the fully-irreducible 4-point vertex, and $\Phi^{c,(N)}$ and $\Gamma^{4,(N)}_c = \Gamma^{4,(N)} - \Phi^{c,(N)}$ represent the 4-point reducible and irreducible vertices in channel $c \in \{ph, \overline{ph}, pp\}$ evaluated to order $N$, respectively. The reducible vertices are defined by
\begin{widetext}
    \begin{flalign}
        \Phi^{ph,(N)}_{\mathbf{k}_1, \mathbf{k}_2; \mathbf{k}_3, \mathbf{k}_4} = \sum_{i+j+k+l=N-1} \Gamma^{4,(i)}_{\mathbf{k}_1, \mathbf{p} + \mathbf{k}_3 - \mathbf{k}_1; \mathbf{k}_3, \mathbf{p}} \Gamma^{4,ph,(j)}_{\mathbf{p}, \mathbf{k}_2; \mathbf{p} + \mathbf{k}_3 - \mathbf{k}_1, \mathbf{k}_4} G^{(k)}_{\mathbf{p}} G^{(l)}_{\mathbf{p} + \mathbf{k}_3 - \mathbf{k}_1}, \label{eq:phi_ph}                               \\
        \Phi^{\overline{ph},(N)}_{\mathbf{k}_1, \mathbf{k}_2; \mathbf{k}_3, \mathbf{k}_4} = \zeta\sum_{i+j+k+l=N-1} \Gamma^{4,(i)}_{\mathbf{k}_1, \mathbf{p} + \mathbf{k}_2 - \mathbf{k}_3; \mathbf{k}_3, \mathbf{p}} \Gamma^{4,\overline{ph},(j)}_{\mathbf{p}, \mathbf{k}_2; \mathbf{p} + \mathbf{k}_2 - \mathbf{k}_3, \mathbf{k}_4} G^{(k)}_{\mathbf{p}} G^{(l)}_{\mathbf{p} + \mathbf{k}_2 - \mathbf{k}_3}, \label{eq:phi_phbar} \\
        \Phi^{pp,(N)}_{\mathbf{k}_1, \mathbf{k}_2; \mathbf{k}_3, \mathbf{k}_4} = \frac{1}{2}\sum_{i+j+k+l=N-1} \Gamma^{4,(i)}_{\mathbf{k}_1, \mathbf{k}_2; \mathbf{p}, \mathbf{k}_1 + \mathbf{k}_2 - \mathbf{p}} \Gamma^{4,pp,(j)}_{\mathbf{p}, \mathbf{k}_1 + \mathbf{k}_2 - \mathbf{p}; \mathbf{k}_3, \mathbf{k}_4} G^{(k)}_{\mathbf{p}} G^{(l)}_{\mathbf{k}_1 + \mathbf{k}_2 - \mathbf{p}}, \label{eq:phi_pp}
    \end{flalign}
    \begin{figure}
        \includegraphics[width=0.97\columnwidth]{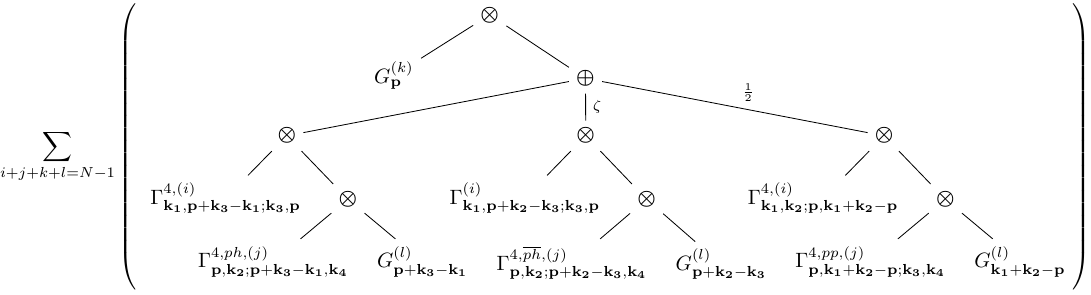}
        \caption{The compact computational graph representation of the the $N$th-order four-point vertex minus fully-irreducible four-point vertex $\Gamma^{4,(N)}_{\mathbf{k}_1, \mathbf{k}_2; \mathbf{k}_3, \mathbf{k}_4} - I^{(N)}_{\mathbf{k}_1, \mathbf{k}_2; \mathbf{k}_3, \mathbf{k}_4}$, based on Eqs.~\eqref{eq:phi_ph}--\eqref{eq:phi_pp}.
        }
        \label{fig:parquet_graph}
    \end{figure}
\end{widetext}
where $\zeta = -1$ for fermions, and $+1$ for bosons, and are shown diagrammatically in Fig.~\ref{fig:phi_diags}. The fully-irreducible 4-point vertex $I^{(N)}$ is built bottom-up from the bare interaction and lower-order Green's functions and 4-point vertex, as shown in Fig.~\ref{fig:envelope_diag}.

The perturbative expansion of Eq.~\eqref{eq:parquet_eqns} leads to the formation of a computational graph as shown in Fig.~\ref{fig:parquet_graph}. The intermediate nodes in this graph represent distinct groups of 4-point vertex function sub-diagrams that share the same order and set of internal variables but differ in topology. Through these intermediate nodes, the computational graph constructs higher-order vertex function diagrams in a hierarchical, self-similar structure. This organization enables efficient reuse of intermediate nodes in constructing higher-order elements, thereby reducing computational redundancy.

\subsubsection{Bottom-up construction of compact computational graphs}

\begin{figure}
    \centering
    \includegraphics[width=1.0\linewidth]{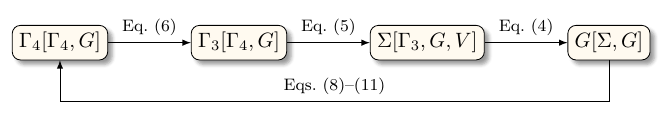}
    \caption{Schematic representation of the bottom-up construction process for compact computational graphs. The cycle shows how Feynman diagrams of different orders are systematically generated through Eqs.~\eqref{eq:Dyson}--\eqref{eq:phi_pp}. Each transformation (arrow) represents specific equations generating new diagrams, with brackets containing the lower-order components required for construction. This cyclic process enables efficient reuse of common subgraphs in generating higher-order diagrams.}
    \label{fig:graph_construct}
\end{figure}

The construction of compact computational graphs arises from coupling DSEs with parquet equations.
As illustrated in the computational graph representations in Figs.~\ref{fig:green}--\ref{fig:parquet_graph}, this approach builds higher-order diagrams by recursively combining lower-order subgraphs.
Figure~\ref{fig:graph_construct} presents the bottom-up construction process, where each current-order Feynman diagram is constructed using relevant lower-order components enclosed in brackets.

Beginning with the zeroth-order $\Gamma^4$ generated by the bare interaction $V$, the construction proceeds through a series of transformations defined by Eqs.~\eqref{eq:Dyson}--\eqref{eq:phi_pp}. Each transformation generates specific diagram types: $\Gamma^4[\Gamma^4,G]$ constructs 4-point vertices using previously generated 4-point vertices and Green's functions through the parquet equations~\eqref{eq:parquet_eqns}--\eqref{eq:phi_pp}; $\Gamma^3[\Gamma^4,G]$ builds 3-point vertices; $\Sigma[\Gamma^3,G,V]$ generates self-energy diagrams; and $G[\Sigma,G]$ produces dressed Green's functions from self-energy insertions. This cyclic process enables the hierarchical generation of higher-order diagrams with efficient reuse of common subgraphs.

\begin{figure*}
    \centering
    \begin{minipage}[b]{0.75\textwidth}
        \includegraphics[width=\linewidth]{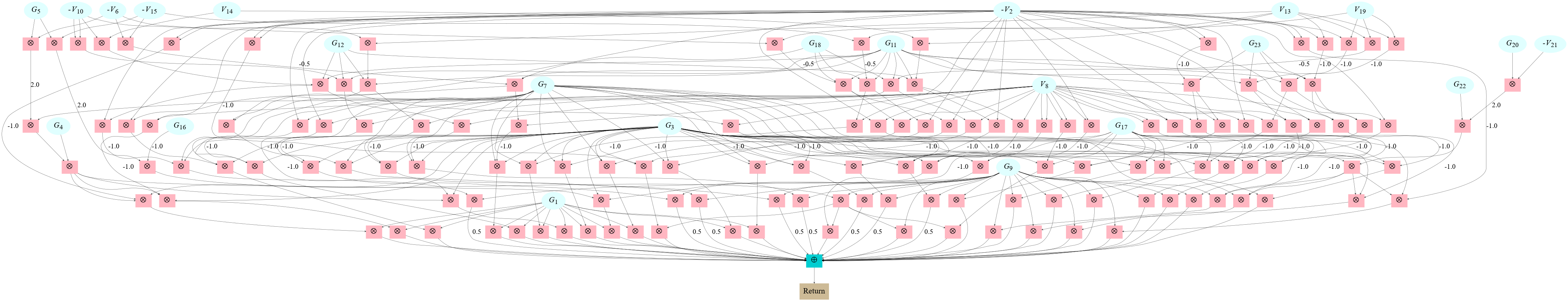}
        \textbf{(a)}
        \label{fig:sigma_flat}
    \end{minipage}
    \hfill
    \begin{minipage}[b]{0.24\textwidth}
        \includegraphics[width=\linewidth]{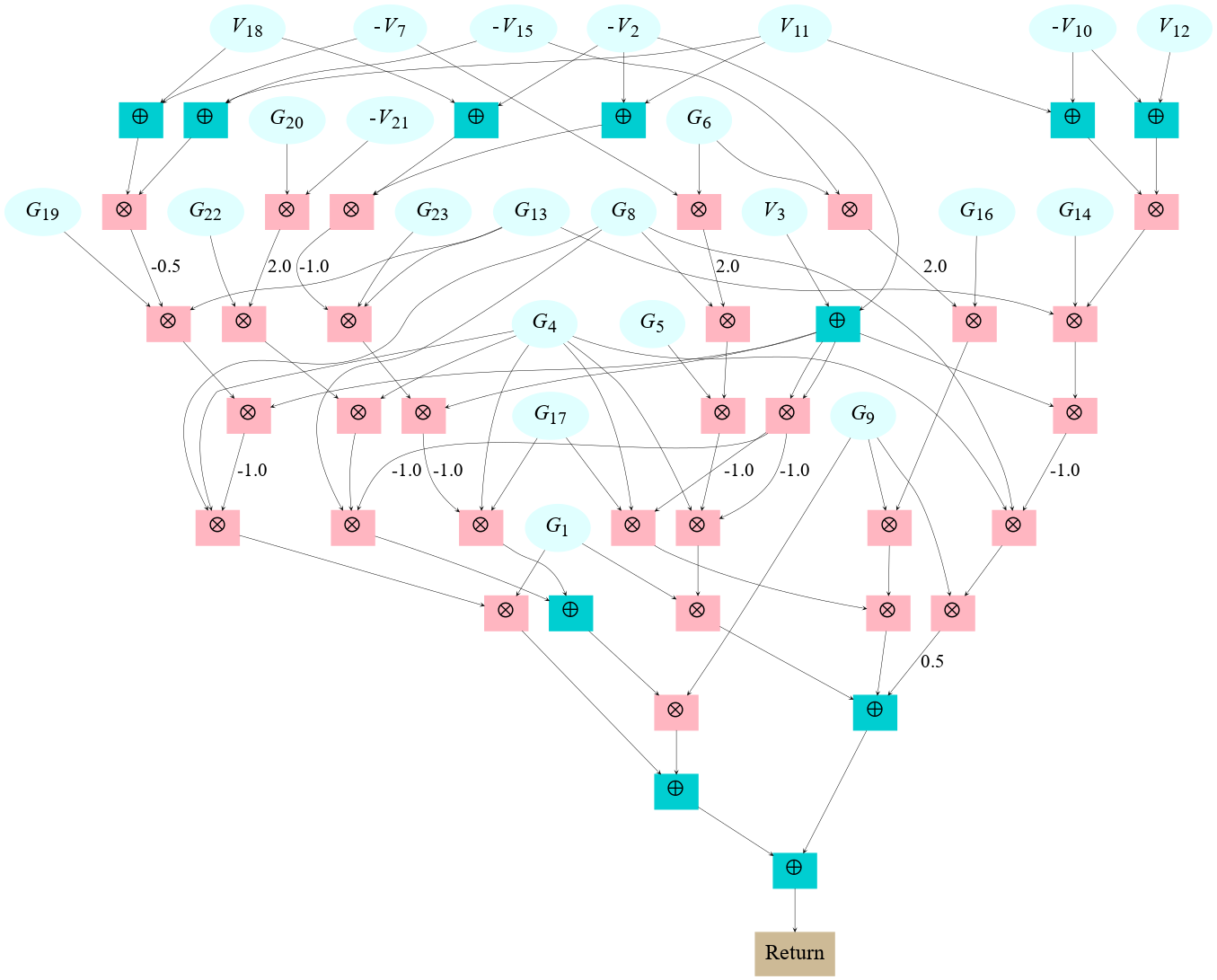}
        \textbf{(b)}
        \label{fig:sigma_parquet}
    \end{minipage}

    \caption{Computational graphs representing third-order Feynman diagrams for the dynamic self-energy of spinless fermions. (a) The graph generated by naive aggregation of Feynman diagrams, with each diagram being a product of propagators `$G$' and interactions `$V$'.
        (b) The optimized graph, derived from the perturbative parquet and Dyson-Schwinger equations, implements an efficient factorization of common sub-diagrams to reduce redundant calculations.
    }
    \label{fig:sigma_graph}
\end{figure*}

\begin{figure}
    \centering
    \includegraphics[width=0.9\columnwidth]{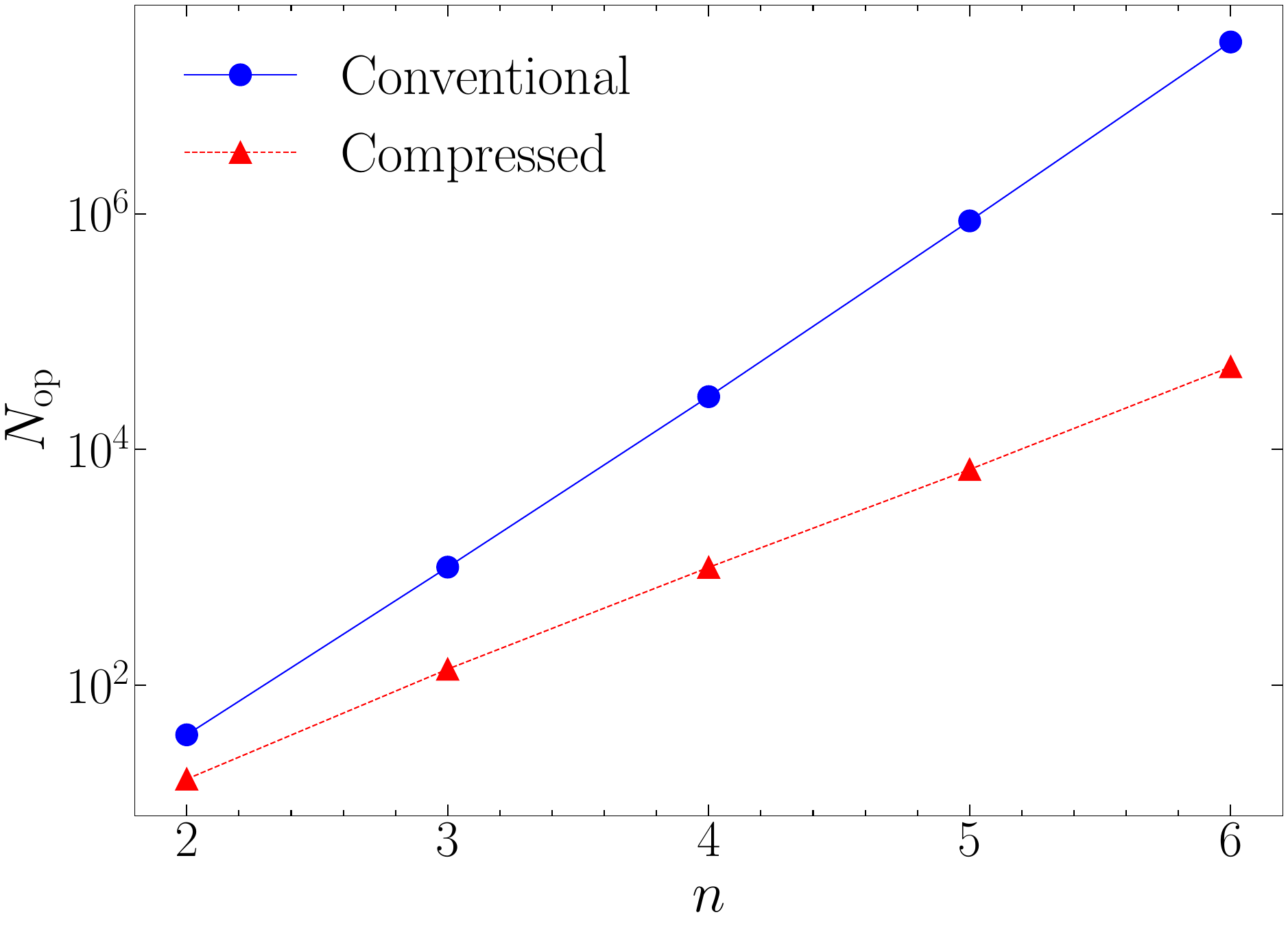}
    \caption{The total number of operations $N_{\rm op}$ required for calculating the $n$th-order self-energy, which is a measure of graph compactness. We compare $N_{\rm op}$ for a traditional summation of Feynman diagrams with the graph generated by Eqs.~\eqref{eq:sigma_pq}--\eqref{eq:phi_pp}.
        The compact computational graph is about three orders of magnitude more efficient than the conventional graph at the 6th order in perturbation theory.}
    \label{fig:parquet_benchmark}
\end{figure}

The effectiveness of this approach is demonstrated in Fig.~\ref{fig:sigma_graph}, which shows the computational graphs for third-order dynamic self-energy diagrams of spinless fermions. Comparing naive aggregation of Feynman diagrams (Fig.~\ref{fig:sigma_graph}(a)) with our optimized graph (Fig.~\ref{fig:sigma_graph}(b)) reveals how factorization of common sub-diagrams significantly reduces redundant calculations. This compression becomes particularly important at higher orders: Figure~\ref{fig:parquet_benchmark} quantifies the total number of operations $N_{\rm op}$ required for calculating the $n$th-order self-energy, comparing traditional Feynman diagram summation against our compressed graph approach. At sixth order, the compact computational graph reduces computational complexity by nearly three orders of magnitude.

The benefits of this bottom-up construction extend beyond minimizing the number of arithmetic operations. Because the Dyson-Schwinger and parquet equations inherently respect the underlying symmetries of the theory, such as crossing symmetry and conservation laws, our construction algorithm provides an automated solution to the challenge of internal loop variable assignment mentioned previously. It implicitly generates diagrammatic groups with the same favorable sign-cancellation properties identified through the explicit symmetry analysis in Ref.~\cite{PhysRevB.103.115141}. Therefore, our framework not only achieves computational efficiency through algebraic factorization but also systematically produces representations that are well-suited to managing the fermionic sign problem.

\section{Renormalization as Automatic Differentiation}
\label{sec:renormalization}

The evaluation of high-order Feynman diagrammatic series, even when represented by optimized computational graphs as discussed in Sec.~\ref{sec:computgraph}, often requires renormalization to yield physically meaningful and convergent results, particularly in many-electron systems. Renormalization systematically reorganizes the perturbative expansion to incorporate low-energy physical scales, thereby improving its predictive power. This section introduces a methodology that combines a bottom-up, field-theoretic ``constructive renormalization scheme" with Taylor-mode automatic differentiation (AD)---a sophisticated technique for exact and efficient evaluation of high-order derivatives---applied directly to the computational graph representation of Feynman diagrams. This synergy enables the efficient and systematic generation of renormalized diagrammatic series to high orders, demonstrating the effective application of AI-driven computational tools to complex problems in QFT.

\subsection{Field-theoretic renormalization}
\label{subsec:rg_background}

Renormalization, initially formulated to address ultraviolet (UV) divergences in relativistic quantum field theories, such as quantum electrodynamics (QED)~\cite{dyson}, provides a general framework for incorporating multiple energy scales in quantum many-body problems.
Wilson's renormalization group paradigm reveals how physical properties depend on the energy scale of observation, yielding fundamental insights into critical phenomena and effective field theories~\cite{wilson1,wilson2}.
Direct computational application of Wilsonian RG to high-order diagrammatic calculations remains computationally demanding. Traditional Dyson-style field-theoretic renormalization methods, therefore, are practically important.
The adaptation of these methods to non-relativistic quantum many-body field theories, where UV divergences are typically absent but large-parameter issues related to short-length scales necessitate similar resummations, represents an active research area~\cite{renorm1,shifted_action,renorm3,kchen,Rossi_2020,Kim_2021homotopic}.

In field-theoretic perturbation theory, renormalization involves a redefinition of bare parameters (associated with a high-energy cutoff or unphysical starting point) into renormalized parameters that reflect the relevant low-energy physics. For instance, a bare propagator $\hat{g}$ is transformed into a renormalized Green's function $\hat{G}$, effectively shifting the starting point of the perturbation. This is achieved by partitioning the original action $S$ into a renormalized part $S_R$ and a counterterm part $S_{CT}$:

\begin{equation}
    S \equiv S_R + S_{CT}
\end{equation}
$S_R$ is formulated using renormalized quantities; for example,
\begin{equation}
    \label{eq:renorm}
    S_R = \int_{\mathbf{k}\tau} \bar{\psi}_{\mathbf{k}\tau}\hat{G}^{-1}_{\mathbf{k}\tau}\psi_{\mathbf{k}\tau} + \int_{\mathbf{k}\mathbf{k}'\mathbf{q}\tau} V_\mathbf{q} \bar{\psi}_{\mathbf{k}+\mathbf{q}\tau}\bar{\psi}_{\mathbf{k}'-\mathbf{q} \tau}\psi_{\mathbf{k}' \tau}\psi_{\mathbf{k} \tau}.
\end{equation}
The diagrammatic expansion generated from $S_R$ begins with zeroth-order terms capturing essential low-energy physics. Higher-order contributions from $S_{CT}$ then systematically cancel any remaining unphysical effects from the bare theory, ensuring the renormalized series reproduces the correct physical results while often exhibiting substantially improved convergence. The specific definitions of $\hat{G}$ and other renormalized parameters constitute a renormalization scheme, whose selection critically affects computational accuracy and efficiency.

\subsection{Constructive renormalization scheme}
\label{subsec:rg_bottomup}

Field-theoretic renormalization can be implemented through top-down or bottom-up approaches. The top-down BPHZ scheme~\cite{bphz1,bphz2,bphz3,bphz4} analyzes individual Feynman diagrams of the bare theory to identify and subtract divergent sub-structures. While foundational, particularly for UV-divergent theories like QED~\cite{QED1,QED2,QED3,fine_constant_theory,fine_const_exp}, its diagram-by-diagram approach becomes
computationally intensive at high orders. Recent developments, such as organizing renormalized diagrams into determinants~\cite{Rossi_2020}, offer significant computational advantages but are typically optimized for real space-time representations, limiting their applicability to momentum-space QFT formalisms.

We employ a bottom-up ``constructive renormalization scheme" particularly suited to quantum many-body problems without UV divergences.
The foundational principles of this scheme are well-established and have proven effective (e.g., in studies of the uniform electron gas~\cite{kchen, dynamic}). Here, we provide a systematic formulation designed to enable the direct application of AD techniques. This three-step procedural framework, integrated with the computational graph representations of Sec.~\ref{sec:computgraph}, manages the complexity of high-order renormalized calculations through:

\begin{enumerate}
    \item \textbf{Re-expansion of the Bare Propagator:} The bare propagator is re-expanded into a power series of the renormalized propagator. This is driven by shifting one or more bare parameters to renormalized parameters defined at the low-energy limit.

          For concreteness, we consider a Fermi liquid system where the bare chemical potential $\mu$ is renormalized to the physically observed chemical potential $\mu_R = E_{\rm F}$, i.e., the Fermi energy. The bare propagator is then re-expanded as
          \begin{equation}
              g(\mu) = G(\mu_R) +  \delta \mu \frac{\partial G(\mu_R)}{\partial\mu_R} +  \frac{\delta \mu^2}{2 !}\frac{\partial^2 G(\mu_R)}{\partial\mu_R^2} + \ldots ,
          \end{equation}
          where the renormalized propagator $G(\mu_R)$ is set as $g(\mu_R)$. While the chemical potential shift $\delta \mu = \mu-\mu_R$ is known, we expand it as a power series in the chemical-potential counterterms,
          \begin{equation}
              \delta \mu = \sum_i \xi^i \delta_{\mu}^{(i)},
          \end{equation}
          where $\delta_{\mu}^{(i)}$ represents the $i$-th order counterterm to be determined by renormalization conditions, and $\xi$ is an auxiliary parameter introduced to track the total perturbation order, including both interactions and counterterms. Note that  $\delta_{\mu}^{(0)}=0$ since the shift is interaction-driven. After enumerating all the diagrams of a given order, we set $\xi = 1$.

    \item \textbf{Construction of Renormalized Feynman Diagrams:} The renormalized Feynman diagrams, as well as their counter-diagrams, are constructed from conventional Feynman rules using the re-expanded propagators. The diagrammatic series results in a double expansion in terms of both the interaction strength and the chemical potential shift. For instance, the shifted self-energy of the system can be organized in a double power series,
          \begin{equation}
              \Sigma(\xi, \delta \mu) = \sum_{n,m} \xi^n\frac{\delta \mu^m}{m!} \Sigma^{(n,m)}(\mu_R), \label{eq:sigma_xi_dmu}
          \end{equation}
          where $\Sigma^{(n,m)}(\mu_R) \equiv \frac{\partial^m \Sigma^{(n)}}{\partial \mu_R^m}$ represents the contribution from self-energy diagrams with $n$ interaction lines and $m$ chemical potential counterterms.

    \item \textbf{Imposing Renormalization Conditions:} The final step is to determine the chemical-potential counterterms $\delta_\mu^{(i)}$ by matching the theory with measured quantities, known as the renormalization conditions. Since the renormalized chemical potential $\mu_R$ is set as the physical Fermi energy, the chemical potential renormalization from the shifted self-energy must vanish at the Fermi surface (FS).
          At each perturbative order $\xi^n$, this condition reads:
          \begin{equation}
              \left.\sum_{b=0}^{n-1} \sum_{\{b_1, b_2, \ldots, b_{n-1} \}}  \prod_i \frac{ \left( \delta_{\mu}^{(i)} \right)^{b_i}}{b_{i}!} \Sigma^{(n-b, m)} (\mu_R) \right|_{\rm FS} \equiv 0 \,,
              \label{eq:counterterm}
          \end{equation}
          where the second summation spans all solutions in nonnegative
          integers $b_i$ satisfying $b_1+2 b_2+\cdots+(n-1) b_{n-1}=b$ and $ b_1+ \cdots+b_{n-1}= m$.
\end{enumerate}

The effectiveness of this renormalization scheme depends on the efficient computation of the Taylor coefficients $\Sigma^{(n,m)}$, which are essentially (functional) derivatives of the original self-energy Feynman diagrams. Note that in typical diagrammatic Monte Carlo calculations, the counterterm order $m$ can be as high as $n \sim 6$. As discussed previously, constructing a compact computational graph for the coefficients $\Sigma^{(n,0)}$ is straightforward, but the high-order derivatives for $m>0$ introduce complexity. In the next subsection, we discuss how to derive a compact representation that includes these high-order derivatives.

Alternative approaches exist to improve series convergence~\cite{shifted_action,Kim_2021homotopic}, diverging from the traditional field-theoretical renormalization schemes. While sharing the common goal of enhancing series convergence, these methods yield distinct mathematical characteristics. It is worth noting that the techniques developed in the subsequent subsection are adaptable to these alternative approaches.

\begin{table}
    \includegraphics[width=\columnwidth]{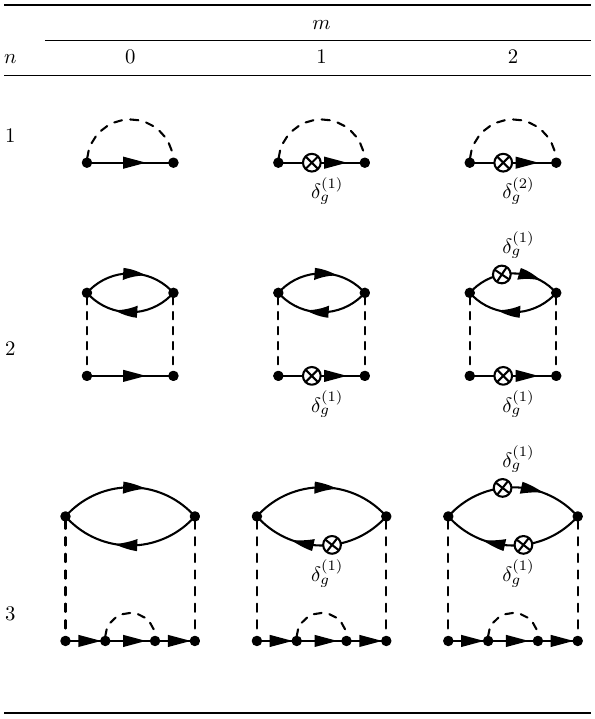}
    \caption{Examples of self-energy diagrams entering Eq.~\eqref{eq:sigma_xi_dmu} grouped by number of interaction lines $n$ and chemical potential counterterm order $m$. The renormalized propagator $G(\mu_R)$ and bare interaction $V$ are represented by directed and dashed lines, respectively, and the Green's function counterterms $\delta^{(n)}_g$ are defined in Eq.~\eqref{eq:green_ct}.}
    \label{tab:sigma_diags_renorm}
\end{table}

\subsection{Taylor-mode automatic differentiation}
\label{subsec:taylormode}

The high-order derivative terms $\Sigma^{(n,m)}$ (Eq.\eqref{eq:sigma_xi_dmu}), essential for the constructive renormalization approach outlined in Sec.\ref{subsec:rg_bottomup}, present a significant computational challenge. We address this by deploying Taylor-mode AD---a sophisticated technique drawn from the modern AI technology stack---directly onto the computational graph representation of Feynman diagrams. Unlike symbolic manipulation or finite differencing, AD computes exact derivatives by systematically decomposing functions and applying the chain rule. Taylor-mode AD is specifically designed for the efficient recursive computation of high-order derivatives of complex, multi-variable functions~\cite{engel_optimal_2023,zhouyin_automatic_2023,Taylor2}, making it well-suited for the constructive renormalization scheme.

Our adaptation of Taylor-mode AD to renormalized Feynman diagrams operates on two main principles:

\begin{enumerate}
    \item \textbf{Truncated Taylor Series of Nodes:} Each node in the computational graph, after renormalization, becomes a truncated Taylor series in the renormalization parameter $\mu_R$. For a given node $f(\mu_R)$, this is represented as:
          \begin{equation}
              \label{eq:taylor_node}
              f(\mu_R) = \sum_{m=0}^{N-1} \frac{\delta \mu_R^m}{m!}f^{(m)}(\mu_R) + \mathcal{O}(\delta \mu_R^N),
          \end{equation}
          where $f^{(m)}$ denotes a compact computational graph for the sub-diagrams with $m$ counterterms, and $N$ is bounded by the truncated diagrammatic order.

    \item \textbf{Composite Function Derivatives:} When a node $f(\mu_R)$ feeds into a higher-level node $g(f(\mu_R))$, forming a composite function, the $m$th-order derivative of $g(\mu_R)$ is dictated by Faà di Bruno's formula:
          \begin{equation}
              \label{eq:chain_rule}
              g^{(m)}(\mu_R)= \sum \frac{m !}{b_{1} ! b_{2} ! \cdots b_{m} !} g^{(k)}(f) \prod_{l=1}^m
              \left(\frac{f^{(l)}(\mu_R)}{l !}\right)^{b_l},
          \end{equation}
          where the sum spans all solutions in nonnegative integers $b_i$ satisfying $b_1+2 b_2+\cdots+m b_m=m$ and $k:=b_1+\cdots+b_m$.
\end{enumerate}

Applying these equations to different operators defines universal chain rules that can be recursively applied to construct a compact computational graph for high-order derivatives. For simple arithmetic operators like addition and multiplication, the chain rules are straightforward. For more complex operators, the chain rules can be automatically generated by a systematic algorithm provided in Ref.~\cite{Taylor3}.

Starting from leaf nodes with known Taylor series (e.g., the renormalized propagator $G(\mu_R)$), these rules are applied recursively to construct a compact computational graph yielding all required $\Sigma^{(n,m)}$ terms simultaneously. This framework is robust to complex graph structures, including those emerging from renormalizing nonlocal QFT interactions.

\begin{figure}
    \centering
    \includegraphics[width=0.9\columnwidth]{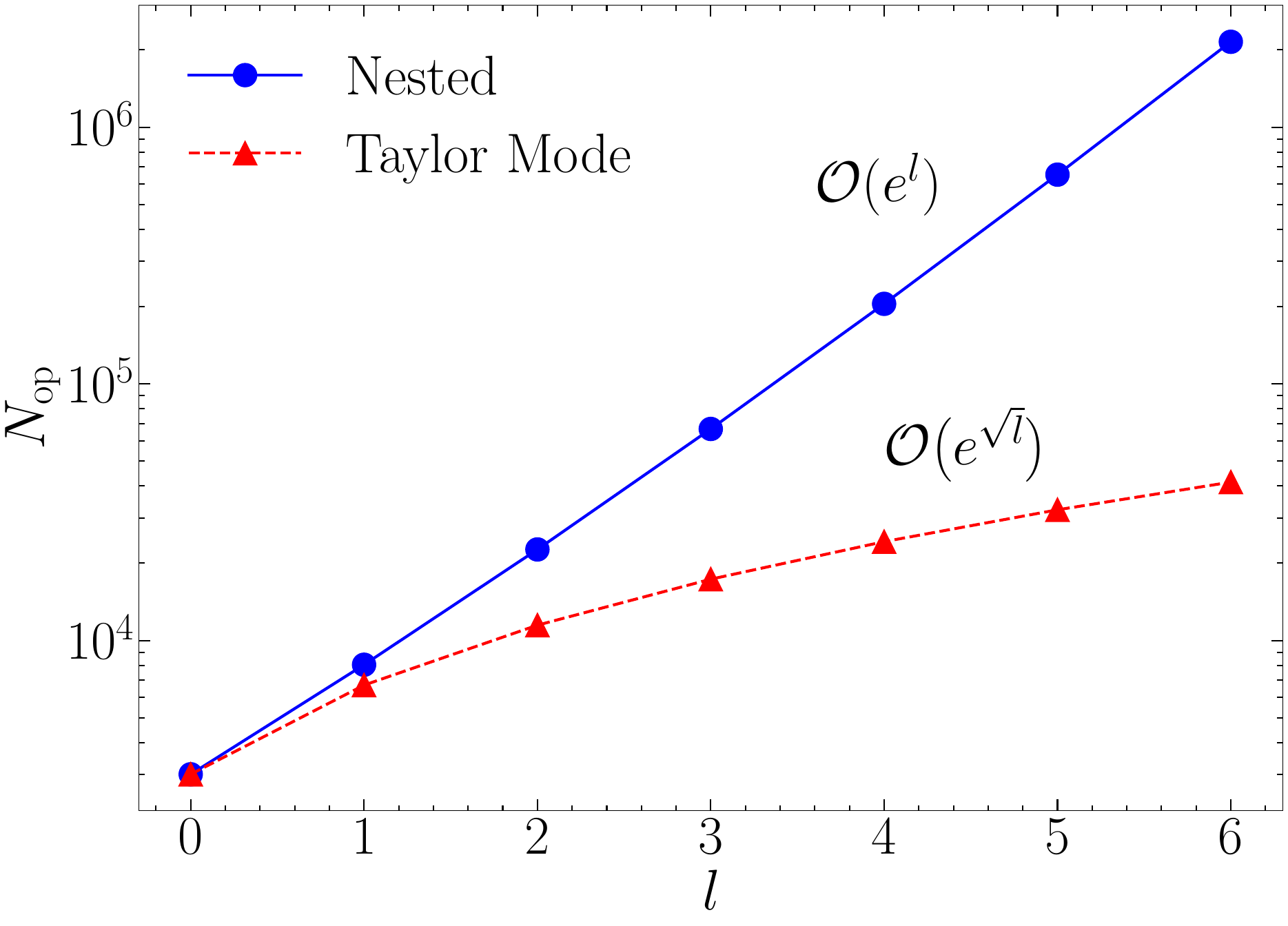}
    \caption{
        Operation counts $N_{\rm op}$ in computational graphs for $l$th-order interaction counterterms of all 4th-order $(n=4)$ self-energy diagrams generated by Eqs.\eqref{eq:sigma_pq}--\eqref{eq:phi_pp}.
        The graphs are generated by differentiating with respect to $\lambda_R$ up to 6th order in $l$ (see Eq.~\eqref{eq:interaction_ct}).
        Nested first-order AD yields $N_{\rm op}$ scaling as $\mathcal {O}(e^{l})$ with $l$, while Taylor-mode AD reduces this to $\mathcal {O}\big(e^{\sqrt{l}}\big)$.}
    \label{fig:ADscaling}
\end{figure}

Taylor-mode AD provides superior computational scaling for high-order derivatives compared to repeated application of first-order AD. By systematically managing the combinatorial explosion (often related to Bell polynomial evaluations intrinsic to high-order chain rules), it mitigates exponential operation growth. This leads to significantly more favorable scaling dependent on the graph topology. Figure~\ref{fig:ADscaling} demonstrates this advantage for $l$-th order interaction counterterms from 4th-order self-energy diagrams generated by Eqs.~\eqref{eq:sigma_pq}--\eqref{eq:phi_pp}, where Taylor-mode AD reduces complexity from $\mathcal{O}(e^{l})$ to $\mathcal{O}(e^{\sqrt{l}})$. This efficiency gain underscores the suitability of Taylor-mode AD for demanding high-order diagrammatic QFT calculations.

\section{Monte Carlo Integration with Neural-network Importance Sampling}
\label{sec:highD_MC}

Evaluating high-order Feynman diagrams in QFT involves computing integrals of the generic form $I_f = \int d\mathbf{x} f(\mathbf{x})$, where $f(\mathbf{x})$ is the diagram's integrand and $\mathbf{x}$ represents the high-dimensional vector of integration variables (e.g., internal momenta and imaginary times). For many-electron systems with Fermi surfaces, $f(\mathbf{x})$ is typically characterized by sharp singularities and complex structures. The widely used VEGAS algorithm~\cite{lepage_new_1978,lepage_adaptive_2021}, while effective for some QFT problems, often struggles in this regime because its separable importance sampling map ($q(\mathbf{x}) = \prod_{k=1}^{D} q_{k}(x_k)$) inadequately captures the strong correlations and singularities induced by Fermi surfaces, resulting in inefficient sampling and high variance. To address this, DiagMC methods employ MCMC techniques. Although MCMC can sample from general distributions, its efficiency critically depends on the proposal mechanism and may be hindered by slow convergence if the mechanism lacks sufficient adaptivity.

Recent advances in the AI technology stack, specifically normalizing flows (NFs), provide powerful tools for learning complex probability distributions~\cite{kobyzev_normalizing_2021}. NFs construct expressive, invertible transformations $T: \mathbf{z} \mapsto \mathbf{x}$ from a simple base distribution $p_z(\mathbf{z})$ (e.g., Gaussian) to a target $p_x(\mathbf{x})$. The probability density of the transformed variable $\mathbf{x}$, given the neural network parameters $\bm{\phi}$ of the flow, is obtained via the change of variables formula:
\begin{equation}
    p_x(\mathbf{x})=p_z(T^{-1}(\mathbf x;\bm \phi))|\det J_{T^{-1}}(\mathbf x;\bm \phi)| \,,
    \label{eq:NFtransform}
\end{equation}
where $J_{T^{-1}}(\mathbf x;\bm \phi)$ is the Jacobian matrix of the inverse transformation $T^{-1}$ with respect to $\mathbf{x}$ and $\bm \phi$ represents the parameters of the NF model. Although NFs have been explored for direct importance sampling in Monte Carlo contexts~\cite{muller_neural_2019,gao_i-_2020,PhysRevD.101.076002,normalization_flow}, their integration into adaptive MCMC frameworks for challenging QFT integrals remains less developed.

We introduce an NF-MCMC methodology that combines the expressive power of NFs with the robustness of MCMC sampling, particularly to tackle singular Feynman diagram integrals in many-electron problems. Our approach utilizes NFs, specifically neural spline flows (NSFs)\cite{durkan2019cubic,durkan_neural_2019} built from coupling layers (see Appendix~\ref{appendix:A}), to construct a highly adaptive proposal distribution $q(\mathbf{x})$ that replaces and significantly outperforms the VEGAS map. This NF-generated $q(\mathbf{x})$ (derived from $|\det J_{T^{-1}}|^{-1}$) serves as the intelligent core of a Metropolis-Hastings MCMC algorithm. The MCMC framework samples configuration points $\mathbf{x}_i$ from a target distribution $\pi(\mathbf{x})$:
\begin{equation}
    \pi(\mathbf{x}) = \alpha q(\mathbf{x}) + (1-\alpha) p(\mathbf{x}),
    \label{eq:target_distr_nf_mcmc_qft_revised}
\end{equation}
where $p(\mathbf{x}) \equiv |f(\mathbf{x})|$ is the absolute integrand, and $\alpha \in [0,1]$ is a mixing parameter. The Metropolis-Hastings step, with acceptance probability $R = \min \left\{ 1, [\pi(\mathbf{x}) Q(\mathbf{x}_t|\mathbf{x})] / [\pi(\mathbf{x}_t) Q(\mathbf{x}|\mathbf{x}_t)] \right\}$ (where $Q$ is the proposal probability, typically $q(\mathbf{x})$ for independent NF proposals), acts to correct any residual imperfections in the NF-learned $q(\mathbf{x})$, ensuring convergence to the true distribution even when the NF imperfectly models $p(\mathbf{x})$. Optional random-walk proposals in the base space of the NF $\mathbf{z}$ can further enhance ergodicity~\cite{hackett_flow-based_2021}.
After discarding initial thermalization steps and taking measurements at appropriate intervals to reduce autocorrelation, we collect $N_{\text{eval}}$ measurements to estimate the integral:
\begin{equation}
    I_f \approx \left.\sum_{i=1}^{N_{\text{eval}}} \frac{f(\mathbf{x}_i)}{\pi(\mathbf{x}_i)} \,\middle/\, \sum_{i=1}^{N_{\text{eval}}} \frac{q(\mathbf{x}_i)}{\pi(\mathbf{x}_i)}\right., \quad \text{where }  \mathbf{x}_i \sim \pi(\mathbf x).
    \label{eq:nf_mcmc_estimator_qft_final_revised}
\end{equation}
This estimator provides the integral value through normalization of $q(\mathbf{x})$ without directly integrating $\pi(\mathbf{x}_i)$, where $\mathbf{x}_i$ represents the set of measured configurations from the equilibrated Markov chain.

This NF-MCMC method provides an efficient framework for evaluating high-dimensional complex integrals
while maintaining statistical reliability. Compared to traditional DiagMC methods, it provides a learned, highly adaptive proposal mechanism that better captures integrand structure. Compared to direct NF importance sampling, it incorporates MCMC's corrective capabilities to ensure convergence even when the NF imperfectly models the target distribution.

\begin{figure}
    \centering
    \includegraphics[width=0.9\linewidth]{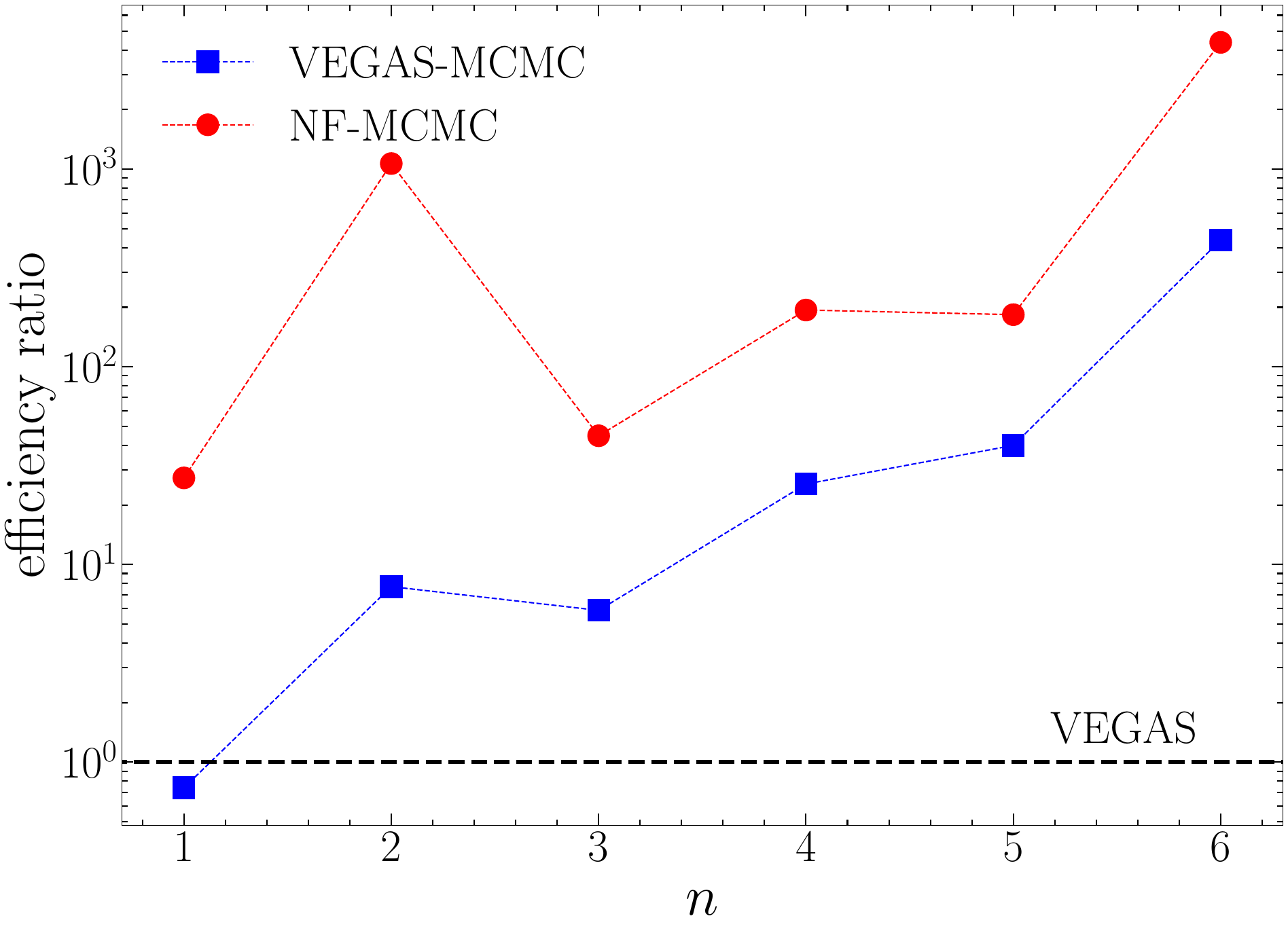} 
    \caption{
        Relative efficiency of  VEGAS-Markov chain Monte Carlo (VEGAS-MCMC) and normalizing flow-MCMC (NF-MCMC) compared to standard VEGAS for the $n$th-order static self-energy at the Fermi surface in the uniform electron gas with $T / T_F = 1 / 32$, where $T_F$ is the Fermi temperature.
        Both MCMC approaches implement the framework of Eqs.~\eqref{eq:target_distr_nf_mcmc_qft_revised}--\eqref{eq:nf_mcmc_estimator_qft_final_revised}, with the mixing parameter $\alpha$ optimized at each order to minimize statistical error.
        Both NF-MCMC and VEGAS-MCMC show significant advantages over standard VEGAS, particularly for higher-order diagrams. NF-MCMC further demonstrates a consistent improvement over VEGAS-MCMC for all observed orders.
        The reduced efficiency at the third order reflects the emergence of the fermionic sign problem.}
    \label{fig:benchmark_MC_nf_mcmc_qft_revised_v2}
\end{figure}

Figure~\ref{fig:benchmark_MC_nf_mcmc_qft_revised_v2} validates our methodology by comparing its efficiency with the standard VEGAS importance sampling and an intermediate VEGAS-MCMC approach. VEGAS-MCMC serves as a crucial benchmark: it employs the same MCMC sampling strategy (Eqs.~\eqref{eq:target_distr_nf_mcmc_qft_revised}--\eqref{eq:nf_mcmc_estimator_qft_final_revised}) but uses the conventional adaptive VEGAS map to generate $q(\mathbf{x})$, rather than the NSF model used in NF-MCMC.
The results are plotted as efficiency ratios relative to standard VEGAS (dashed line at 1) for $n$th-order static self-energy diagrams of the UEG, with the mixing parameter $\alpha$ optimized for each diagram order to minimize statistical error.

The VEGAS-MCMC approach (blue squares) outperforms standard VEGAS for $n > 1$, demonstrating that the MCMC framework with $q(\mathbf{x})$ can improve upon traditional importance sampling by correcting imperfections in the VEGAS map.
The NF-MCMC approach (red circles) achieves significantly higher efficiency than both VEGAS and VEGAS-MCMC across all diagram orders. This demonstrates the substantial benefit of using a more expressive NF-generated proposal $q(\mathbf{x})$, which better captures the complex structure of the Feynman integrand, resulting in improved acceptance rates and reduced variance within the MCMC framework. Efficiency gains are particularly pronounced for higher-order diagrams, where integrand complexity increasingly challenges simpler sampling schemes.

A notable feature is the decrease in relative efficiency for NF-MCMC and VEGAS-MCMC at $n=3$. This is characteristic of encountering the fermionic sign problem, where the integrand $f(\mathbf{x})$ develops strongly oscillating positive and negative regions.
This complicates the NF's ability to learn an optimal representation of $|f(\mathbf{x})|$ and limits potential efficiency gains for any MCMC method that relies on integrand magnitude. Despite this challenge, the overall trend favors NF-MCMC, demonstrating its superior capability for complex QFT integrations.

\section{AI Tech Stack for Many-Electron Field Theory}
\label{sec:AI4QFT_implementation_final}

Our overall methodology is implemented through an integrated computational framework designed as a tailored AI technology stack for QFT calculations. This framework is built upon two principal, synergistic components that address distinct but related computational challenges. First, we present a specialized compiler for the systematic translation of Feynman diagrams into optimized computational graphs and the integrated execution of field-theoretic renormalization. Second, we describe a heterogeneous computing architecture for the efficient calculation of high-dimensional Monte Carlo integrations, particularly deploying the NF-MCMC scheme of Sec.~\ref{sec:highD_MC} on massively parallel hardware such as GPUs.
The unifying principle across both components is the representation of diagrammatic calculations as computational graphs,
enabling the deployment of advanced algorithms and optimization techniques across diverse computing platforms.

\subsection{Feynman diagram compiler with integrated automatic differentiation}

\begin{figure}
    \centering
    \includegraphics[width=0.9\linewidth]{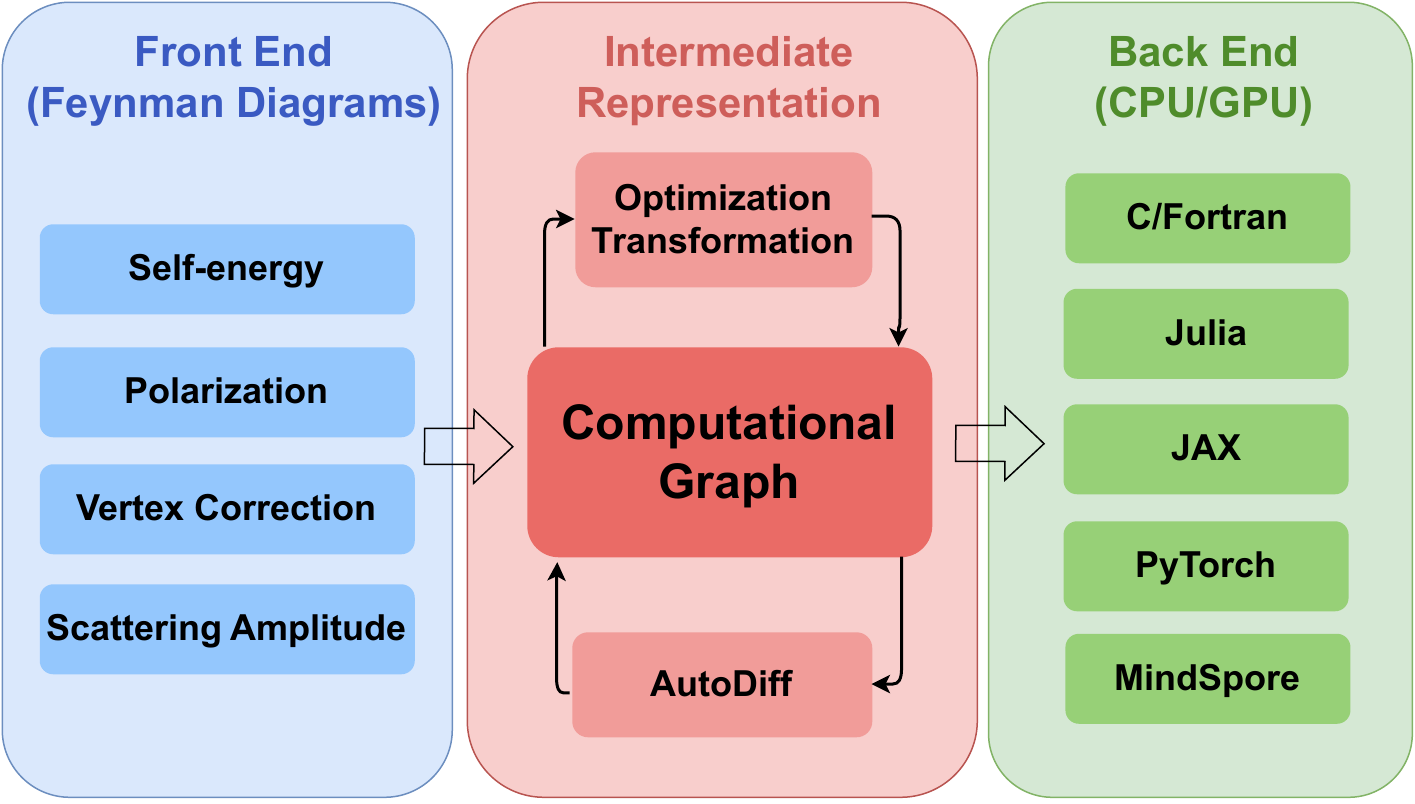} 
    \caption{Architecture of the Feynman diagram compiler. Feynman diagrams are translated into a unified computational graph (Intermediate Representation, IR). This IR enables local optimizations and the application of Taylor-mode automatic differentiation for renormalization. The back-end generates optimized code for target platforms, including CPUs and GPUs.}
    \label{fig:compiler_prx_final}
\end{figure}

The core of our diagrammatic machinery is a compiler (Fig.~\ref{fig:compiler_prx_final}) that translates abstract Feynman diagrams into executable code. This process mirrors the architecture of modern programming language compilers.

The compiler operates through a three-stage pipeline. The Front-End parses Feynman diagrams and converts them into a unified Intermediate Representation (IR) in the form of a computational graph. In this graph, the leaf nodes typically represent fundamental QFT entities (propagators, vertex interactions),  providing the initial symbolic or numerical values for calculations.
Intermediate nodes represent mathematical operations (addition, multiplication, exponentiation) that combine inputs and other intermediate results. Edges denote the flow of intermediate calculations, which ultimately lead to the overall value of the Feynman integrand.

Subsequently, the IR processing stage performs crucial transformations on the static computational graph. Key operations include graph optimization, where local rewriting rules are applied to simplify expressions and reduce redundant computations, and renormalization via Taylor-mode AD. As detailed in Sec.\ref{subsec:taylormode}, Taylor-mode AD is applied directly to the graph IR to compute the necessary derivative terms $\Sigma^{(n,m)}$.
This static graph representation for AD facilitates deeper optimizations compared to dynamic AD approaches, which is particularly beneficial for the complex high-order derivatives inherent in QFT.

Finally, the Back-End translates the optimized and renormalized graph IR into high-performance source code.
Although the compiler itself is developed in Julia~\cite{Julia-2017}, its back-end can flexibly generate code for various targets, including native CPU code and code compatible with ML frameworks such as JAX~\cite{jax2018github}, TensorFlow~\cite{tensorflow2015-whitepaper}, and PyTorch~\cite{NEURIPS2019_9015}.
This cross-platform capability is essential for integration with the existing AI tech stack.
The compiler is available as an open-source Julia package: \emph{FeynmanDiagram.jl}~\cite{FeynmanDiagram_2023}.

\begin{figure}
    \centering
    \includegraphics[width=0.85\linewidth]{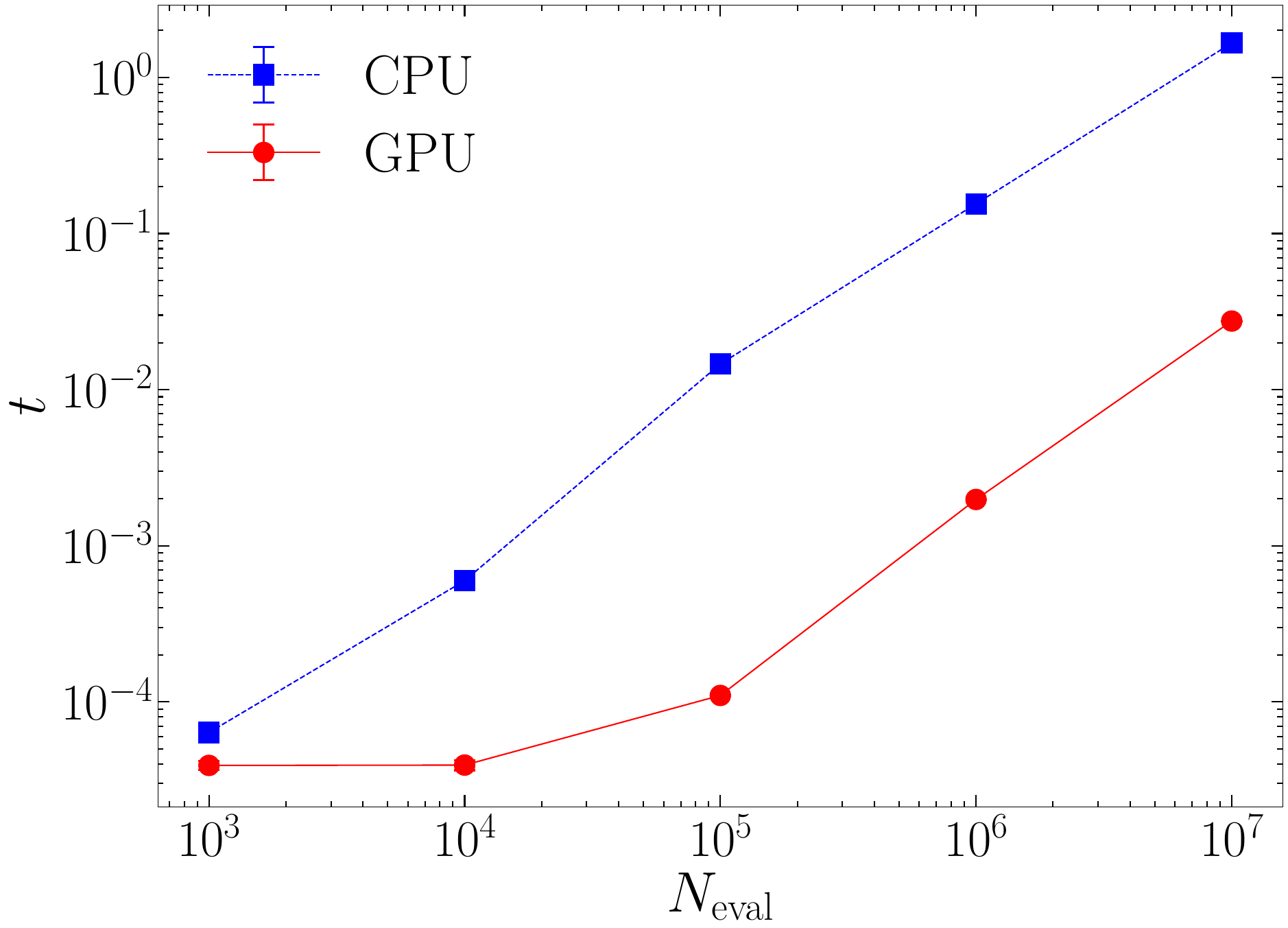}
    \caption{Evaluation time for a 4th-order self-energy computational graph versus batch size $N_{\text{eval}}$. A JAX implementation on an A100 GPU achieves an approximately 100-fold speedup at large batch sizes over a CPU implementation (i7-11700, single-thread), illustrating the computational advantages of GPU parallelization for QFT calculations.
    }
    \label{fig:GPU_prx_final}
\end{figure}

Figure~\ref{fig:GPU_prx_final} illustrates the computational performance gains achievable through GPU parallelization in QFT calculations through benchmarking the evaluation of a 4th-order self-energy diagram's computational graph.
Using a JAX implementation on both CPU and A100 GPU architectures, we measure execution times across varying batch sizes ($N_{\text{eval}}$), where JAX’s vectorized mapping function (vmap) efficiently parallelizes operations across batched inputs.
The performance scaling reveals distinct regimes: at small batch sizes ($N_{\text{eval}} \lesssim 10^4$), GPU acceleration provides only a modest speedup due to insufficient parallelization to compensate for overhead costs and saturate GPU resources.
As $N_{\text{eval}} \gtrsim 10^5$, however, full GPU utilization is achieved, resulting in approximately 100-fold acceleration over CPU execution. Additionally, we find that the JAX CPU implementation outperforms the traditional serialized C implementation, likely attributed to JAX's utilization of vector support features present in modern CPU architectures.

Future work may explore tensor network contraction algorithms~\cite{tensor1, tensor2, tensor3, tensor4} for further optimizations, particularly for diagrams with complex tensor structures. The integration of compact computational graph representations with Taylor-mode AD distinguishes our approach from other efforts applying ML to diagrammatic computations, such as Ref.~\cite{burke_torchami_2023}.

\subsection{GPU-accelerated NF-MCMC integration via computational graphs}
\label{subsec:GPU-NFMCMC}

The second component of our computational framework is the efficient execution of the NF-MCMC algorithm (Sec.~\ref{sec:highD_MC}) on heterogeneous computing platforms, primarily GPUs. Traditional MCMC methods face challenges in parallelizing effectively on GPUs due to their sequential nature and conditional branching. Our NF-MCMC design inherently addresses these limitations.

\begin{figure}
    \centering
    \includegraphics[width=0.55\linewidth]{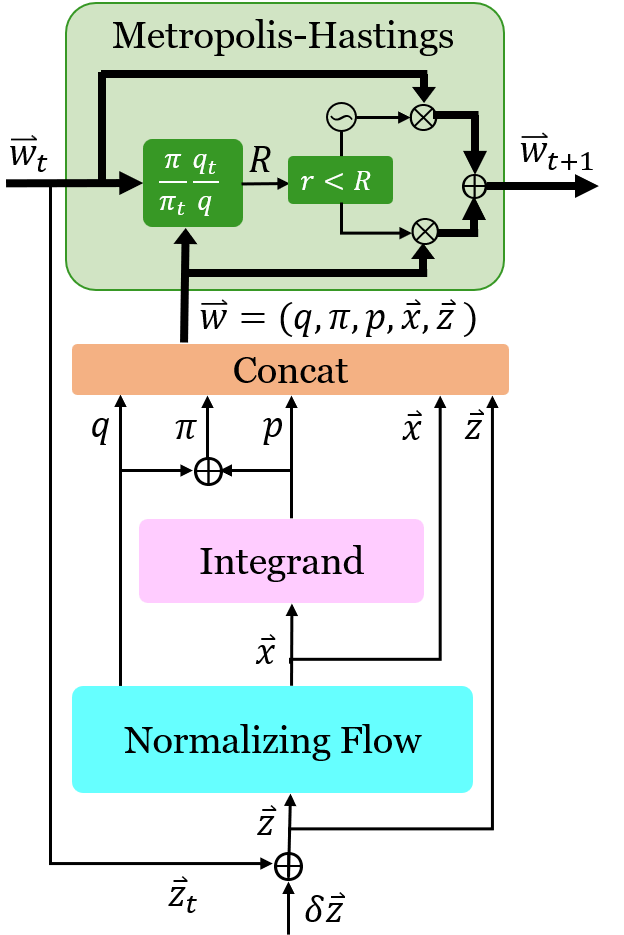} 
    \caption{Computational graph for a single NF-MCMC update step. Variables $\mathbf{w} \equiv (q,\pi,p,\mathbf{x},\mathbf{z})$ flow through sequential operations: normalizing flow mapping $\mathbf{z} \to \mathbf{x}$, integrand calculations for $p(\mathbf{x})$, and Metropolis-Hastings acceptance $R$. This structure minimizes branching, enabling efficient batching of many independent Markov chains on GPUs.}
    \label{fig:NF_MCMC_prx_final}
\end{figure}

As shown in Fig.~\ref{fig:NF_MCMC_prx_final}, the entire NF-MCMC update step is structured as a single, largely branch-free computational graph. This graph encompasses proposal generation via the NF transformation $T: \mathbf{z} \mapsto \mathbf{x}$, evaluation of the QFT integrand $f(\mathbf{x})$ (which itself constitutes a computational graph generated by our compiler), calculation of the relevant probability densities $q(\mathbf{x})$ and $p(\mathbf{x})$, and the Metropolis-Hastings acceptance decision. The concatenated variables $\mathbf{w} \equiv (q,\pi,p,\mathbf x,\mathbf z)$ represent the data flowing sequentially through these operations. The only significant branch arises from the acceptance check ($r < R$, where $r$ is a uniform random number), a condition efficiently managed by modern GPU hardware.

This graph-based representation of the MCMC logic is key to unlocking GPU parallelism. It allows us to run many independent Markov chains simultaneously through batched computational graphs. This batching is efficiently implemented using vectorization functionalities provided by ML frameworks, such as JAX's vmap function, which maps the single-chain graph computation across a batch of chain states. Such an approach ensures high arithmetic intensity and promotes regular memory access patterns, both essential for optimal GPU performance.

Furthermore, the integrand evaluation component $f(\mathbf{x})$, generated by our Feynman diagram compiler as an optimized JAX-compatible graph, seamlessly integrates into the larger NF-MCMC graph. This creates a fully GPU-accelerated pipeline, from abstract diagram definition to numerical integration, significantly boosting computational throughput for high-dimensional QFT integrations while preserving the statistical rigor of the MCMC methodology.

\section{Application to the Effective Mass of the Uniform Electron Gas}
\label{sec:meff_example}

In this section, we apply our algorithms and renormalization techniques, as elaborated in Sections~\ref{sec:computgraph}--\ref{sec:highD_MC}, to calculate the effective mass ratio $m^*/m$ in the UEG model.

The quasiparticle effective mass $m^*/m$, a fundamental parameter in Landau Fermi liquid theory, characterizes how electron-electron interactions modify electron mobility~\cite{landau_book, silin1, silin2, FL1, FL2}. This parameter is essential for understanding electron behavior in diverse materials.

Precise determination of $m^*/m$ in the UEG presents substantial challenges due to limitations in current computational techniques. Quantum Monte Carlo methods, including diffusion Monte Carlo (DMC)~\cite{ceperley_ground_1980} and variational Monte Carlo (VMC)~\cite{PhysRevB.16.3081}, provide powerful numerical approaches for treating many-body correlations, but depend heavily on the choice of trial wave functions and are susceptible to finite-size effects. On the other hand, traditional diagrammatic perturbation theory, such as the GW approximation, faces two main constraints: results vary significantly with the chosen renormalization scheme, and calculations are often restricted to low orders, thereby hindering reliable systematic error estimation and result validation~\cite{yasuhara1992effective}. These differing approaches yield not just varying but contradictory results for the effective mass~\cite{simion_many-body_2008,haule2022single, azadi2023, holzmann2023static}.

This lack of methodological consensus for the UEG effective mass behavior underscores a critical gap in our ability to address nonlocal electron interactions, motivating the development of an alternative systematic computational framework.

\subsection{Model and methods}

The UEG model represents interacting electrons in a uniform, inert, neutralizing background\textemdash the quintessential model of an electron liquid. Its simplicity and fundamental relevance make it a cornerstone in the study of electronic structures in materials science. The UEG provides a foundational framework for understanding electronic behavior in a broad spectrum of materials.

The bare action of the UEG is described by:
\begin{flalign}
    S_0 = & \int_{0}^{\beta} d\tau \Bigg( \sum_{\mathbf{k}\sigma}\bar{\psi}_{\mathbf{k}\sigma} (\partial_\tau + \epsilon_k - \mu) \psi_{\mathbf{k}\sigma} \nonumber \\ &+ \frac{1}{2V}\sum_{\substack{\mathbf{q} \ne 0 \\ \mathbf{k}\mathbf{k^\prime}\sigma\sigma^\prime}}  V_q \bar{\psi}_{\mathbf{k+q}\sigma} \bar{\psi}_{\mathbf{k^\prime - q}\sigma} \psi_{\mathbf{k^\prime}\sigma^\prime} \psi_{\mathbf{k}\sigma} \Bigg),\label{eq:ueg_action}
\end{flalign}
where $\psi$ and $\bar{\psi}$ are Grassmann fields, $\epsilon_k = \frac{\hbar^2 k^2}{2m}$  is the kinetic energy term representing the UEG dispersion, and $V_q = \frac{4\pi e^2}{q^2}$ is the Coulomb interaction. Notably, the interaction strength in the UEG model inversely correlates with the electron density, so that an increase in density leads to weaker interactions. The density $\rho$ of the UEG is parameterized by the (dimensionless) Wigner-Seitz radius $r_s \equiv \bar{r} /a_0 $, where $\bar{r} = (4\pi \rho/3)^{-1/3}$ is the average interparticle distance, and $a_0$ is the Bohr radius.

We now demonstrate how our method enables precise high-order perturbative calculations of the effective mass of the UEG.
The central quantity in this problem is the momentum-frequency resolved self-energy $\Sigma(k, i\omega)$, from which the quasiparticle properties are extracted. The effective mass ratio $m^*/m$ is given by
\begin{flalign}\label{eq:eff_mass_ratio}
    \frac{m^*}{m} = Z^{-1} \cdot
    \left(1 +
    \frac{m}{k_{\rm F}}\left.\frac{\partial \text{Re}\Sigma(k, 0)}{\partial k}\right|_{k = k_{\rm F}}\right)^{-1},
\end{flalign}
where the renormalization constant
\begin{flalign}
    Z = \left(1 - \left.\frac{\partial\text{Im}\Sigma(k_{\rm F}, i\omega)}{\partial \omega}\right|_{\omega = 0}\right)^{-1}
\end{flalign}
gives the strength of the quasiparticle pole.
We note that the form of $Z$ arises from the analytic properties of the self-energy in the fermionic Matsubara formalism, since the derivative of the (even) real part, $\text{Re}\Sigma$, vanishes at $\omega=0$.
While $Z$ is a ground-state property, the notation $|_{\omega=0}$ signifies the zero-frequency limit, which we practically obtain by extrapolating the finite-difference expression involving the lowest Matsubara frequency to $T\to 0$. The accurate determination of the effective mass thus depends entirely on a robust calculation of the self-energy.

Our approach employs a diagrammatic method to calculate the self-energy at a given temperature $T$. A direct perturbative treatment based on the bare Coulomb interaction is, however, problematic, leading to an unphysical divergence of the effective mass at the one-loop level, even in the weakly interacting regime. This necessitates a robust renormalization scheme for a physically meaningful perturbative series.

We adopt the renormalized field theory approach of Ref.~\cite{kchen}, wherein the perturbative expansion is performed using a theory of electrons interacting via an effective, statically screened Yukawa potential. This framework incorporates two crucial renormalization conditions: (1) the chemical potential is adjusted to maintain the physical Fermi energy, and (2) the bare Coulomb interaction is replaced by a statically screened Yukawa interaction. 
This systematically removes large, unphysical parameters from the bare diagrams, dramatically improving the convergence of the diagrammatic series. While such series may ultimately be asymptotic, analogous to perturbative QED, a reorganized series with effectively small coupling parameters remains a powerful predictive tool. Indeed, its application to the UEG has enabled the accurate calculations of various quasiparticle properties up to fifth- or sixth-order DiagMC computations~\cite{haule2022single, ko, wdm, dynamic}, demonstrating practical efficacy.

Here, we provide a reformulation of this approach compatible with the Taylor-mode AD. Following the renormalization procedure of Sec.~\ref{sec:renormalization}, we re-expand the UEG action (Eq.~\eqref{eq:ueg_action}) in terms of a renormalized propagator $G(\mu_R)$ and interaction $V_q(\lambda_R)$, which are characterized by a renormalized chemical potential $\mu_R$ and static screening parameter $\lambda_R$, respectively, yielding the counterterms
\begin{flalign}
    g(\mu) - G({\mu_R})    & = \sum^\infty_{n=1} \frac{\delta\mu^n}{n!} \frac{\partial^n G(\mu_R)}{\partial\mu^n_R} = \sum^\infty_{n=1} \delta^{(n)}_{g}, \label{eq:green_ct} \\[1ex]
    V_q - V_q({\lambda_R}) & = \sum^\infty_{n=1} \frac{\delta\lambda^n}{n!} \frac{\partial^n V_q({\lambda_R})}{\partial\lambda^n_R} = \sum^\infty_{n=1} \delta^{(n)}_{V}.
    \label{eq:interaction_ct}
\end{flalign}
As discussed in Sec.~\ref{subsec:rg_bottomup}, the renormalized chemical potential is set to the physical Fermi energy $\mu_R = E_{\rm F}$. Here $V_q({\lambda_R}) = \frac{4\pi e^2}{q^2 + \lambda_R}$ is a Yukawa interaction\textemdash the bare interaction is recovered when $\lambda_R = 0$. Unlike the post-processing approach to the counterterm series of $\delta\mu$ (Eq.~\eqref{eq:sigma_xi_dmu}), the free parameter $\delta\lambda = -\xi\lambda_R$ is treated as a variational constant to optimize the convergence of the perturbative expansion.

In the present case, the self-energy is thus expanded as a power series in three variables $\xi$, $\delta \mu$, and $\delta \lambda$,
\begin{equation}
    \Sigma(\xi, \delta \mu, \delta \lambda) = \sum_{n,m,l} \xi^{n}\frac{\delta \mu^m}{m!}\frac{\delta \lambda^l}{l!} \Sigma^{(n,m,l)}(\mu_R, \lambda_R), \label{eq:sigma_xi_dmu_dlambda}
\end{equation}
where $\Sigma^{(n,m,l)}(\mu_R, \lambda_R) \equiv \frac{\partial^m \partial^l \Sigma^{(n)}}{\partial \mu_R^m \partial \lambda_R^l}$ is the self-energy contribution from diagrams with $n$ interaction lines, $m$ chemical-potential counterterms, and $l$ interaction counterterms.
The diagrams corresponding to $\Sigma^{(n,m,l)}$ parallel those listed in Tab.~\ref{tab:sigma_diags_renorm}, with the addition of $l$ interaction counterterms.

To evaluate the effective mass ratio, we first use Eq.~\eqref{eq:sigma_xi_dmu_dlambda} to express the frequency and momentum derivatives of the self-energy as power series in $\xi$,
\begin{flalign}
    a(\xi, \delta \mu, \delta\lambda) & = \sum^{\infty}_{n=0} \xi^n a_n(\delta\mu, \delta\lambda), \label{eq:dsigma_depsilon_series}
\end{flalign}
where
\begin{flalign}
    a_n(\delta\mu, \delta\lambda) = \sum_{m,l} \frac{\delta \mu^m}{m!}\frac{\delta \lambda^l}{l!} a^{(n,m,l)}\label{eq:dsigma_depsilon_series_1}
\end{flalign}
with analogous definitions for $b(\xi, \delta\mu, \delta\lambda)$ and $b_n(\delta \mu, \delta \lambda)$, and
\begin{flalign}
    a^{(n,m,l)} & = \frac{m}{k_{\rm F}} \left.\frac{\partial\text{Re}\Sigma^{(n,m,l)}(k, 0)}{\partial k}\right|_{k = k_{\rm F}}, \label{eq:deltam_partitions} \\[1ex]
    b^{(n,m,l)} & = \left.\frac{\partial\text{Im}\Sigma^{(n,m,l)}(k_{\rm F}, i\omega)}{\partial \omega}\right|_{\omega = 0}. \label{eq:deltas_partitions}
\end{flalign}
Taylor-mode AD enables us to compute the momentum derivative in Eq.~\eqref{eq:deltam_partitions} on the fly such that the contributions $a^{(n,m,l)}$ are simulated directly without introducing a discretization error. On the other hand, the frequency derivative in Eq.~\eqref{eq:deltas_partitions} must be estimated by finite difference methods due to the discrete nature of the Matsubara frequency axis at finite temperature $T$, so that $\left. \partial\text{Im}\Sigma(k_F, i\omega)/\partial \omega \right|_{\omega = 0} = \lim_{T \rightarrow 0} \text{Im}\Sigma(k_F, i\pi T) / \pi T$.

\begin{figure}
    \centering
    \includegraphics[width=0.95\columnwidth]{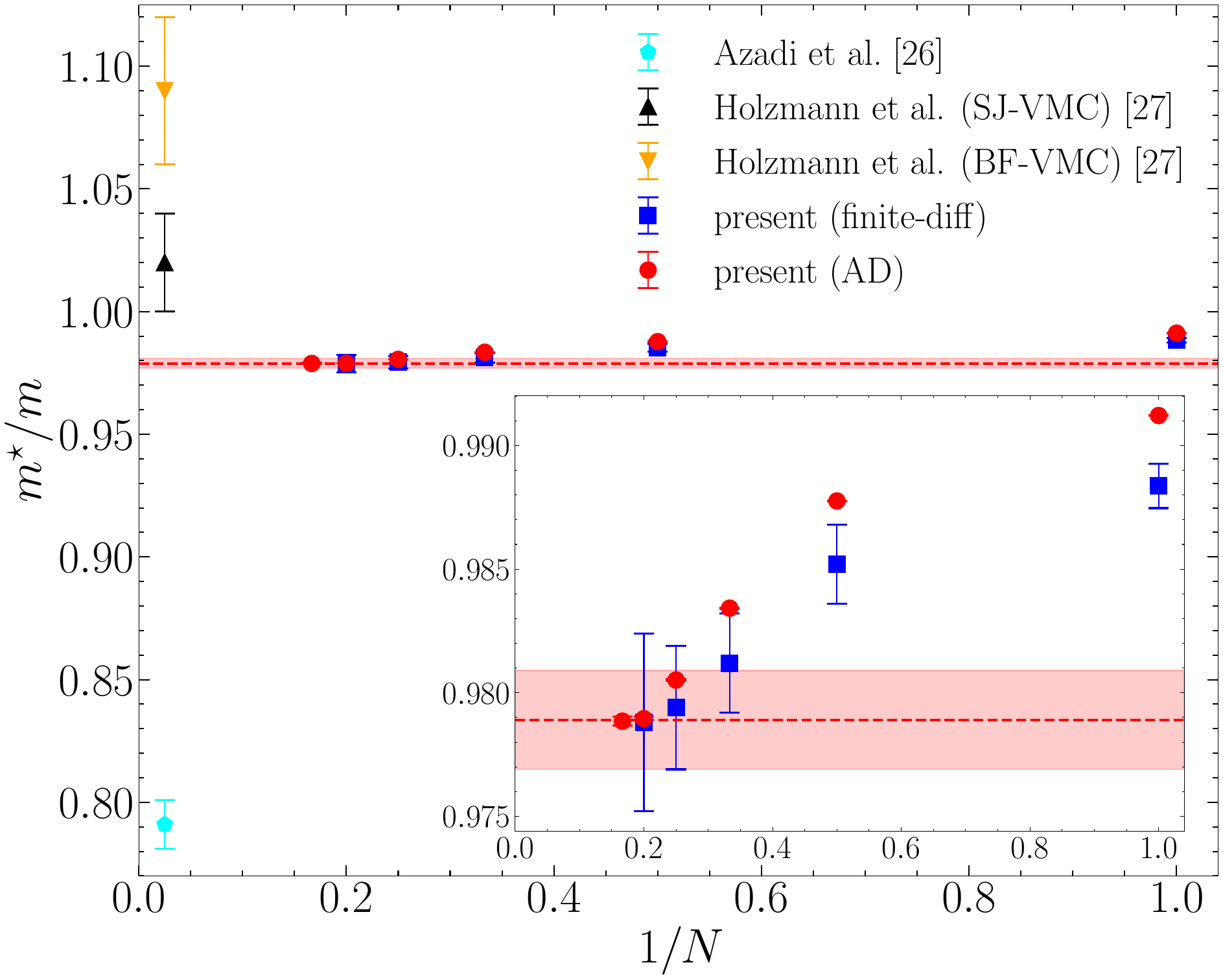}
    \caption{Effective mass ratio $m^*/m$ against perturbation order $N$ for the 3D uniform electron gas at $r_s=5$, $T/T_F=1/40$, and $\lambda_{R}=1.375$. The tuning of $\lambda_R$ ensures optimal numerical convergence without altering the physical outcomes. The graph illustrates DiagMC results using Taylor-mode Automatic Differentiation (red circles) and finite differences (blue squares) for computing the coefficients $a^{(n,m,l)}$ (Eq.~\eqref{eq:deltam_partitions}), with both methods producing consistent data. Notably, our results are consistent with those from the recent Variational Quantum Monte Carlo (VMC) studies~\cite{holzmann2023static}, but offer markedly smaller error bars, signifying a substantial improvement in precision. In contrast, there is a significant divergence from other Quantum Monte Carlo results~\cite{azadi2023}.
    }
    \label{fig:meff_vs_N}
\end{figure}

The computational graph methodology yields a simple and efficient procedure to compute the UEG effective mass ratio to order $N = n+m+l$ in the renormalized perturbation theory, which may be summarized as follows:
\begin{enumerate}
    \item Construct computational graphs for the self-energy (see Fig.~\ref{fig:sigma_graph}) and its high-order derivatives $a^{(n,m,l)}$ and $b^{(n,m,l)}$.

    \item
          Perform DiagMC integration using the constructed computational graphs, employing VEGAS-MCMC for stochastic sampling of momenta and imaginary-time variables. As described in Sec.~\ref{sec:highD_MC}, the variables are proposed following an adaptively learned distribution from the VEGAS map~\cite{lepage_new_1978,lepage_adaptive_2021} and sampled by the target distribution (Eq.~\eqref{eq:target_distr_nf_mcmc_qft_revised}).

    \item Perform the renormalization post-processing procedure to compute the chemical potential shift $\delta \mu$ and the power series coefficients $b_n(\delta\mu, \delta\lambda)$ and $a_n(\delta\mu, \delta\lambda)$.

    \item Compute the effective mass ratio $m^*/m = (1 - b) \cdot (1 + a)^{-1}$ from its power series expansion in $\xi$.
\end{enumerate}
A notable advantage of this approach is that the computational graphs for the self-energy derivative contributions $a^{(n,m,l)}$ and $b^{(n,m,l)}$ can be precompiled once and for all up to a given order $N$ and then reused for Monte Carlo simulations at, for example, different values of the screening parameter $\lambda_R$ or temperature $T$.

Our approach can be compared with the recent connected determinant DiagMC algorithm for the Hubbard model~\cite{iv_two-dimensional_2022}, which implements a double expansion in interaction strength and chemical potential counterterms to achieve numerically exact calculations at fixed particle density. In contrast, the multi-parameter expansion we employ emerges naturally from our framework's ability to automatically compute high-order derivatives, constituting a formal Taylor series of the action around physical reference points. This provides systematic routes to multi-parameter renormalization broadly applicable to any counterterms formulated as action derivatives, rather than being algorithmically tailored to specific models.

Our method for calculating the self-energy and effective mass of the UEG incorporates a significantly greater number of Feynman diagrams compared to Ref.~\cite{haule2022single}. While this comprehensive treatment increases computational demands, it eliminates dependence on dressed interactions parameterized in prior studies~\cite{haule2022single}. This independence enables exploration of previously inaccessible regimes: temperatures effectively approaching the zero-temperature limit and density parameters beyond $r_s = 4$. The removal of constraints from pre-determined parameters allows for a more accurate and thorough investigation of the UEG.

\begin{figure}
    \centering
    \includegraphics[width=0.85\columnwidth]{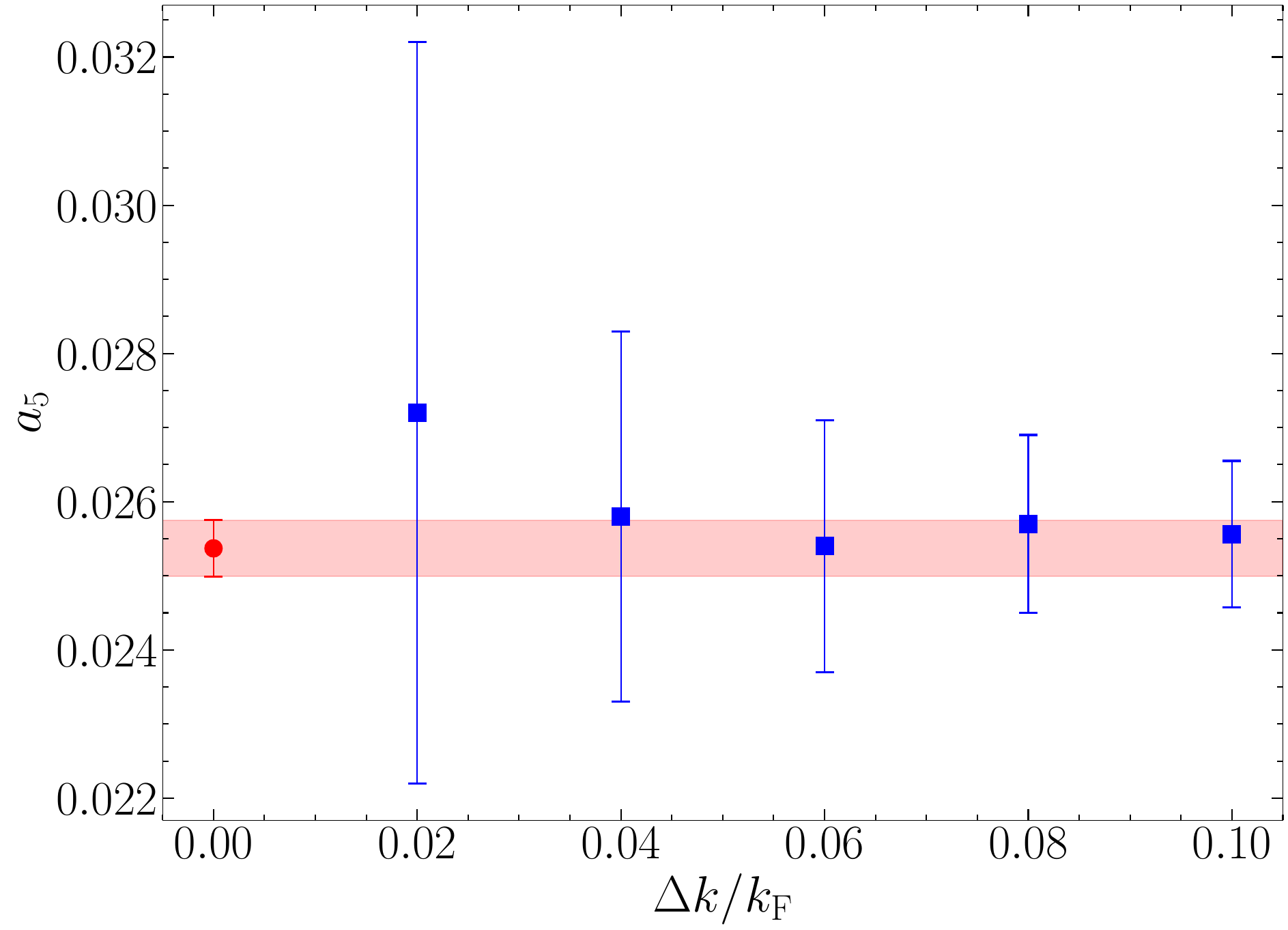}
    \caption{Demonstration of the advantages of AD over numerical finite differentiation in the 3D UEG with $r_s=5$ and $\lambda_R=1.125$. The red circle indicates the AD result for the momentum derivative of the 5th-order self-energy $a_5$.
    The blue squares are the corresponding finite differences $\frac{m}{k_{\rm F}} \frac{\Delta \rm{Re} \Sigma^{(5)}(k,0)}{\Delta k}$ with $\Delta k /k_{\rm F} = k/k_{\rm F}-1 = 0.02, 0.04, 0.06,0.08$ and 0.1. Each data point is sampled for $10^{10}$ Monte Carlo steps.}
    \label{fig:deltam_vs_dk}
\end{figure}

\subsection{Results}

We demonstrate our methodology by calculating the effective mass ratio $m^*/m$ for the 3D UEG at $r_s = 5$. Our high-order diagrammatic calculations at $T/T_F = 1/40$ yield the estimate $m^*/m = 0.979(2)$, as shown in Fig.~\ref{fig:meff_vs_N}. Using an optimized screening parameter of $\lambda_R = 1.375$, we achieve an accurate estimation of $m^*/m$ within our perturbative framework.

Our uncertainty assessment adopts a necessarily heuristic and conservative approach, incorporating both statistical uncertainties from Monte Carlo sampling and systematic uncertainties from perturbative truncation. Given the inherent challenges in assessing the asymptotic behavior of perturbative series, we evaluate systematic uncertainties by analyzing the variation among the last three perturbation orders, as detailed in Appendix~\ref{appendix:B1}. Thermal corrections exhibit linear-$T$ dependence and are approximately an order of magnitude smaller than our total uncertainty at $T/T_F=1/40$ (Appendix~\ref{appendix:B2}). Hence, our result can be directly compared with ground-state effective mass values.

Our result is in quantitative agreement with recent Quantum Monte Carlo (QMC) data from Holzmann et al.~\cite{holzmann2023static}, while improving the reported precision by nearly two orders of magnitude. However, it differs from another QMC results reported by Azadi et al.~\cite{azadi2023}. The most striking physical implication of our result is that the quasiparticle effective mass remains remarkably close to the bare electron mass, even at this strongly correlated density of $r_s=5$. This is particularly noteworthy given that at this density, the average Coulomb potential energy is four times greater than the average kinetic energy, and the quasiparticle weight $Z$ is substantially reduced to below 0.5~\cite{holzmann_momentum_2011}.
The observation that strong correlation effects nearly cancel in their contribution to the effective mass at $r_s=5$  provides important input for understanding correlation effects in electron liquids, which will be addressed in a forthcoming study across different densities and dimensionalities~\cite{mass}.

The AD technique plays a key role in enhancing the computational precision, as demonstrated in the inset of Fig.~\ref{fig:meff_vs_N}. Furthermore, we compare AD with finite difference methods for calculating the momentum derivative of the self-energy $a$ in Fig.~\ref{fig:deltam_vs_dk}. Finite difference methods suffer from systematic errors when using wide spacing $\Delta k$ and from statistical errors when using narrow spacing $\Delta k$. In contrast, AD provides a more robust and error-resistant approach. This improvement in calculating $a$ directly enhances the reliability of the effective mass estimation.

For completeness, we mention that the DiagMC simulation with AD up to the maximum perturbation order $N=6$ shown in Fig.~\ref{fig:meff_vs_N} requires roughly $10^4$ CPU hours on a single-threaded processor. This computational demand, when compared to the resource-intensive nature of traditional methods, underscores the efficiency of our approach, particularly considering the precision and reliability achieved.

\section{Conclusion and Outlook}
\label{sec:conclusion}
We present a comprehensive computational framework that advances the solution of many-electron QFTs by addressing three fundamental numerical challenges. First, we establish a compact computational graph representation for Feynman diagrams with generic algorithms for their construction across diverse physical observables, thereby tackling the complexities of diagram generation and enumeration. Second, by implementing Taylor-mode AD directly on this graph framework, we enable efficient field-theoretic renormalization and the computation of arbitrary (functional) derivatives. Third, we address diagram evaluation through optimized integrand computation via our graph compiler and an NF-MCMC algorithm, specifically designed for massive parallelism on GPUs, to handle high-dimensional integrations.

These advances exploit fundamental parallels between QFT constructs and computational architectures in modern AI technology stacks. The computational graphs and tensor operations inherent to Feynman diagrammatics align naturally with established ML frameworks. This synergy facilitates the systematic deployment of modern ML frameworks~\cite{tensorflow2015-whitepaper,thetheanodevelopmentteam2016theano,NEURIPS2019_9015, jax2018github, mindspore_2023}, which provide optimized environments for computational graph manipulation and efficient execution on diverse hardware platforms, including CPUs, GPUs, and emerging specialized processors.

Embodying this paradigm, we developed a versatile Feynman diagram compiler that translates abstract diagrams into optimized code executable across various computing platforms. Integrating this compiler with our VEGAS-MCMC sampler within a renormalized DiagMC scheme, we computed the effective mass of the three-dimensional UEG model in the strongly correlated regime ($r_s=5$). The renormalization procedure, enabled by Taylor-mode AD, allows systematic treatment of high-order counterterms essential for convergent results.
Our calculations achieve a precision for $m^*/m$ that surpasses state-of-the-art QMC simulations~\cite{holzmann2023static} by nearly two orders of magnitude, demonstrating the quantitative power of our approach.

This framework extends beyond the UEG to applications in first-principles calculations for complex electron liquids in materials, including the determination of advanced exchange-correlation functionals for density functional theory~\cite{Medvedev_2017,Mardirossian_2017,VERMA2020302,wdm,PhysRevB.111.155132}  and Coulomb pseudopotentials for ab initio superconducting transition temperature predictions~\cite{epw_anisotropic_2013,epw_2016,wang_2023}. The approach adapts readily to fundamental quantum many-body models on lattices, such as Hubbard and frustrated spin systems.

Methodologically, our framework opens avenues for implementing and calculating diverse diagrammatic expansions through multiple pathways.
For example, the recent prospect of analytically integrating internal Matsubara frequencies to directly access real-frequency linear response functions~\cite{burke_torchami_2023}, thus bypassing numerical analytic continuation, could benefit significantly from our computational graph approach to optimize Matsubara summation by Taylor mode AD and efficiently evaluate the resulting complex dynamic Feynman diagrams. Similarly, strong-coupling expansions for theories with dominant local interactions~\cite{pairault_strong-coupling_2000,PhysRevB.103.195147}, which generate intricate diagrammatic topologies, represent another fertile ground for exploration with graph-based representations. While our current graph optimizations primarily target scalar operations on CPUs, future developments will incorporate tensor network algorithms~\cite{tensor1, tensor2, tensor3, tensor4} for enhanced performance, particularly on specialized hardware.

In summary, this work establishes a synthesis of QFT methodologies with advanced computational techniques. By systematically integrating the AI technology stack, we enable the efficient computation of previously challenging diagrammatic calculations while offering new capabilities to investigate correlated electron systems. Our demonstrated precision in strongly correlated regimes opens pathways to quantitative many-body theory, establishing foundations for investigating previously intractable systems and advancing theoretical understanding of correlated many-body physics.

\begin{acknowledgments}
    The authors express their gratitude to Xiuzhe (Roger) Luo and Jin-Guo Liu for their engaging and insightful discussions on the Feynman diagram compiler. K.C. extends sincere thanks to Kristjan Haule, Gabriel Kotliar, Nikolay Prokof'ev, Boris Svistunov, Antoine Georges, and Olivier Parcollet for their valuable insights and discussions. The Flatiron Institute is a division of the Simons Foundation. P.H. and Y.D. were supported
    by the National Natural Science Foundation of China
    (under Grant No. 12275263) and the Innovation Program
    for Quantum Science and Technology (under Grant No.
    2021ZD0301900). L.W. is supported by the National Natural Science Foundation of China under Grants No. T2225018, No. 92270107, and No. 12188101, No. T2121001, and the Strategic Priority Research Program of Chinese Academy of Sciences under Grants No. XDB0500000 and No. XDB30000000.
    T.W. and X.C. were supported by the National Science Foundation Grant DMR-2335904.
    D.C. was supported by the US Department of Energy, Office of Basic Energy Sciences as part of the Computation Material Science Program.
    K.C. was supported by National Natural Science Foundation of China under Grants No. 12047503.

\end{acknowledgments}

P.H., T.W., and D.C. contributed equally to this work.

\appendix

\section{Neural Spline Flow for Feynman Diagram Integration}
\label{appendix:A}
Here we provide implementation details of the neural spline flow (NSF) model, which is used in the normalizing flow-Markov chain Monte Carlo (NF-MCMC) algorithm for the integration of high-order Feynman diagrams in Sec.~\ref{sec:highD_MC}. NSF is a specific type of NF that employs piecewise rational-quadratic splines to construct flexible invertible transformations to model complex probability distributions. Our implementation builds on the PyTorch package \emph{normflows}~\cite{Stimper2023}.

\subsection{Normalizing flow architecture and neural spline flow}
NFs are a class of neural
network models that transform a simple base distribution into a more complex target distribution through a
series of invertible and differentiable mappings, offering a flexible framework for constructing tractable sampling distributions. Mathematically, a NF transformation $T$ composed of $K$ mappings can be described as
\begin{equation}
    \mathbf{x} = T(\mathbf{z})= T_K \circ \cdots \circ T_1 (\mathbf{z}) \,,
\end{equation}
where each $T_i$ represents an invertible transformation parametrized by a neural network, and $\mathbf{z}\sim p_z(\mathbf{z})$ is a simple base distribution such as a Gaussian. By this invertible architecture, NFs have the ability to learn expressive probability distributions while maintaining exact density evaluation and efficient sampling. This enables adaptive importance sampling that can capture complex correlations and sharp features in high-dimensional integrands.

A prevalent structure in NFs is the coupling layer, which enables complex transformations while maintaining computational tractability. In a coupling layer, the input variables are split into two disjoint subsets. One subset remains unchanged, while the other is transformed using a function conditioned on the first subset. Mathematically, a coupling transform, which maps a $D$-dimensional input $\mathbf{z}=[\mathbf{z}_{1:d},\mathbf{z}_{d+1:D}]$ to an output $\mathbf{z}^\prime=[\mathbf{z}^\prime_{1:d},\mathbf{z}^\prime_{d+1:D}]$, can be written as:
\begin{equation}
    \begin{aligned}
        \mathbf{z}^\prime_{1:d} & = \mathbf{z}_{1:d}\,,                                          \\
        z^\prime_i              & = g(z_i; \bm{\theta}_i) \quad \text{for $i=d+1,\dots,D$}   \,,
    \end{aligned}
    \label{eq:coupling_layer}
\end{equation}
where $g$ represents a piecewise invertible function conditioned by parameters $\bm{\theta}_{d+1:D}=\text{NN}(\mathbf{z}_{1:d})$, with NN being an arbitrary neural network. The inverse transformation is similarly straightforward, ensuring the invertibility of the flow. By stacking multiple coupling layers, NFs can model highly complex distributions while maintaining the ability to compute exact likelihoods.

NSF represents an advanced implementation within the NF framework, employing piecewise polynomial or rational splines to achieve highly flexible and expressive transformations~\cite{durkan_neural_2019}. These flows partition the input domain into several bins, each defined by a monotonic polynomial or rational segment, providing a smooth mapping that captures complex structures in the target distribution with high precision. Interestingly, the VEGAS map can be viewed as a simple heuristic instance of spline flows, essentially employing zero-order splines with no coupling layers, relying on piecewise constant segments for iterative refinement of the sampling
distribution. In Sec.~\ref{sec:highD_MC}, we propose a NF-enhanced Markov chain Monte Carlo integration method and implement it via an efficient computational graph framework with a GPU-accelerated pipeline (Sec.~\ref{subsec:GPU-NFMCMC}).

\subsection{Implementation of neural spline flow}
\begin{figure}
    \centering
    \includegraphics[width=\linewidth]{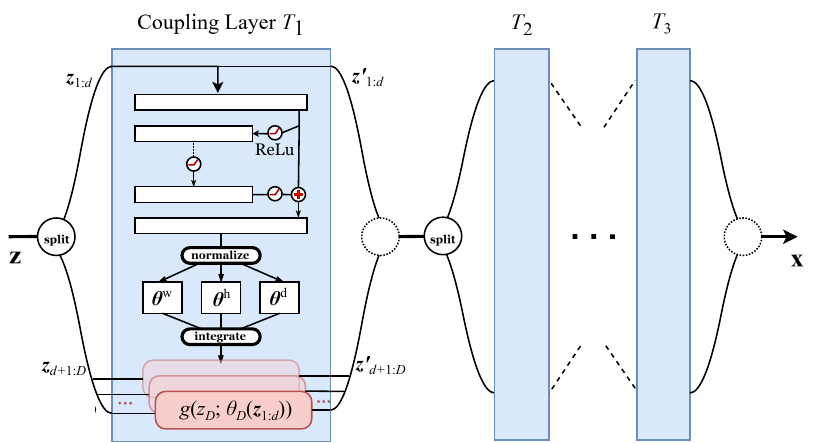}
    \caption{
    The architecture of the coupling layer in the neural spline flow model. The input variables $\mathbf z$ are split into two partitions: $\mathbf z_{1:d}$ (left unchanged) and $\mathbf z_{d-1:D}$ (transformed using a piecewise rational-quadratic spline function $g$ driven by the outputs of a residual neural network with $\mathbf z_{1:d}$ inputs). The outputs are normalized to yield a parameter vector $\mathbf \theta_i=[\mathbf \theta_i^{\rm w}, \mathbf \theta_i^{\rm h}, \mathbf \theta_i^{\rm d}]$. Vectors $\theta_i^{\rm w}$ and $\mathbf \theta_i^{\rm h}$ represent the widths and heights of the spline bins, and $\theta_i^{\rm d}$ represents the derivatives at the internal knots. Multiple coupling layers may be compounded to achieve expressive transformations.
    }
    \label{fig:NSF_layer}
\end{figure}

Key components and settings of the NSF implementation in this work include monotonic rational-quadratic splines for flexible and computationally efficient transformations with 8 spline bins, a residual network (ResNet) with two hidden layers of 32 units each, and ReLU activation functions. The base distribution used is a Uniform distribution. Figure~\ref{fig:NSF_layer} illustrates the structure of the NSF coupling layer, which highlights the use of piecewise coupling transformations and the ResNet architecture.  This design ensures the invertibility of the coupling transform and simplifies inversion, which is crucial for density evaluation and sample generation in Monte Carlo integration.

To address the unique challenges of high-dimensional Feynman diagram integration, we implemented two key techniques to effectively train the NSF model:
\begin{enumerate}
    \item Annealing Training: To handle sharp distributions at low temperatures, we employed an annealing schedule in the temperature parameter. Starting from a high temperature where the distribution is smoother, we gradually decrease it during training, enabling the model to effectively learn the intricate features while ensuring stable convergence.

    \item MCMC Sampling: MCMC methods were integrated to generate samples that accurately represent the complex high-dimensional structures in the Feynman diagrams. This combination allowed the model to produce more effective samples for the training dataset. By combining NSF and MCMC sampling, we achieved a more robust and efficient training process.
\end{enumerate}

The performance of the NSF model was evaluated based on the loss values. The loss is defined by the forward Kullback-Leibler (KL) divergence between the target distribution $p^*_x(\mathbf x)$ and the NSF-learned distribution $p_x(\mathbf x; \bm \phi)$:
\begin{equation}
    \begin{aligned}
        \mathcal{L}(\bm{\phi}) = & D_{\mathrm{KL}}\left[p^*_x(\mathbf{x}) \| p_x(\mathbf{x} ; \bm{\phi})\right]                                                                                            \\
        =                        & -\mathbb{E}_{p^*_x(\mathbf{x})}\left[\log p_{x}(\mathbf{x} ; \bm{\phi})\right]+\text {const.}                                                                           \\
        =                        & -\mathbb{E}_{p^*_x(\mathbf{x})}\left[\log p_z \left(T^{-1}(\mathbf{x} ; \bm{\phi}) \right)+\log \left|\operatorname{det} J_{T^{-1}}(\mathbf x;\bm \phi) \right| \right] \\
                                 & + \text { const. }
    \end{aligned}
\end{equation}
Here, the last equality uses Eq.~\eqref{eq:NFtransform}. By generating a set of Monte Carlo samples $\{\mathbf x_i \}^N_{i=1}$ from $p_x^*(\mathbf x)$, we can estimate the loss value as
\begin{equation}
    \begin{aligned}
        \mathcal{L}(\boldsymbol{\theta}) \approx & -\frac{1}{N} \sum_{i=1}^N \log p_z \left(T^{-1}\left(\mathbf{x}_i ; \boldsymbol{\phi}\right)\right)+\log \left|\operatorname{det} J_{T^{-1}}\left(\mathbf{x}_i ; \boldsymbol{\phi}\right)\right| \\
                                                 & +\text{const.}
    \end{aligned}
\end{equation}
Minimizing this Monte Carlo approximation of the forward KL divergence is equivalent to fitting the NSF model to the samples $\{\mathbf x_i \}^N_{i=1}$ by maximum likelihood estimation.

The final model losses for the first- to sixth-order self-energy diagrams of the uniform electron gas were 0.001, 0.47, 1.0, 1.7, 2.3, and 3.3, respectively.
Combined with the MCMC method, we obtain high-precision integral results, as demonstrated in Sec.~\ref{sec:highD_MC}.  The NSF model's ability to learn adaptive sampling distributions and generate high-quality samples is crucial in our computations, leading to substantial improvements in efficiency and accuracy compared to the conventional numerical integration methods.

\section{Convergence and Uncertainty in the Effective Mass Calculation}
\label{appendix:B}

This appendix provides a detailed analysis of the effective mass calculations for the three-dimensional uniform electron gas at $r_s = 5$ presented in Sec.~\ref{sec:meff_example}. Given the inherent challenges in assessing convergence and systematic uncertainties in high-order perturbative calculations, we present detailed studies of convergence behavior across different screening parameters, systematic uncertainty assessment strategies, and finite-temperature effects. This analysis validates our computational methodology and supports our result $m^*/m=0.979(2)$.

\subsection{Screening parameter optimization and systematic uncertainty assessment}
\label{appendix:B1}

\begin{figure}
    \centering
    \includegraphics[width=0.9\linewidth]{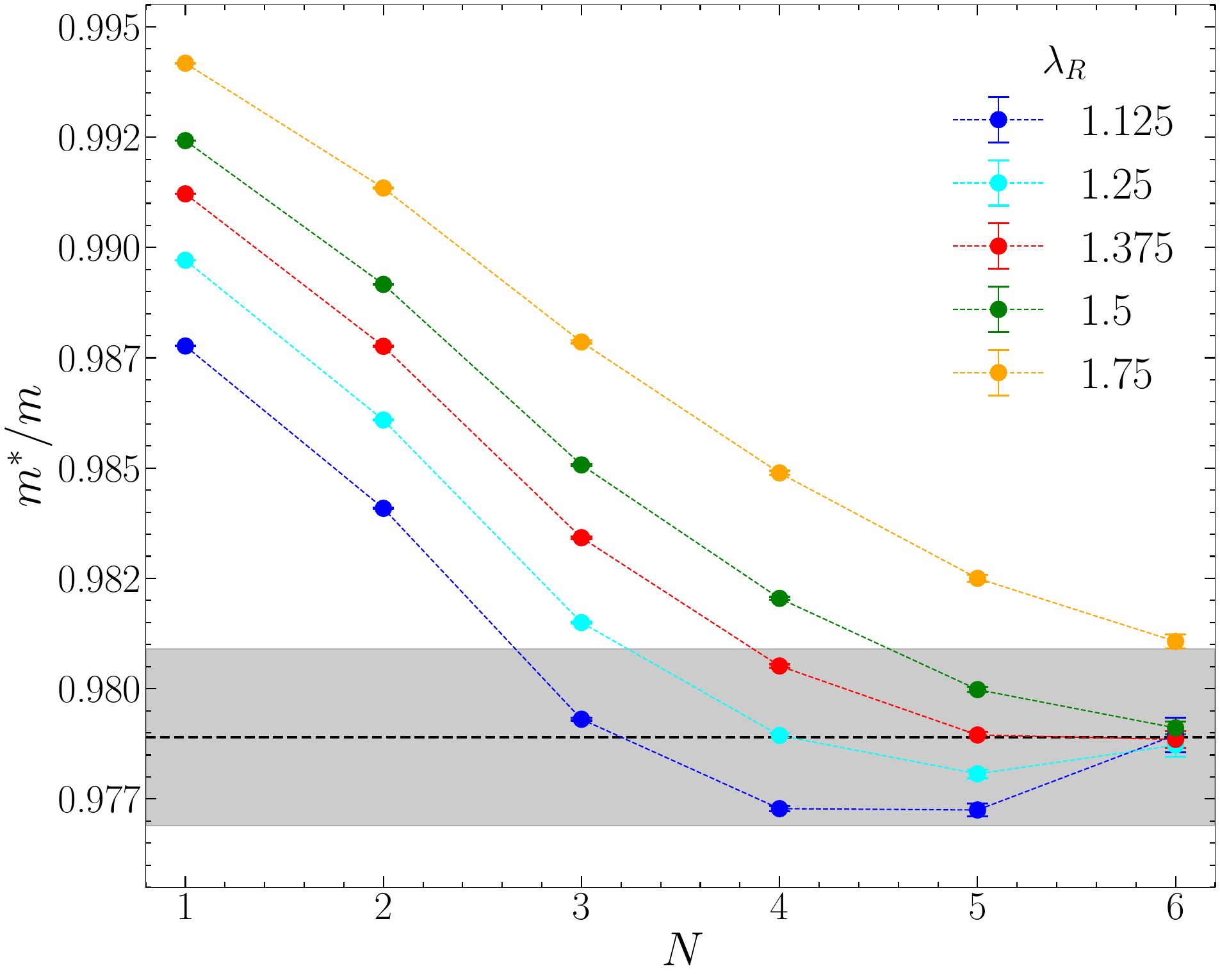}
    \caption{Effective mass ratio $m^*/m$ versus perturbation order $N$ for different screening parameters $\lambda_R$ at $T/T_F = 1/40$ and $r_s = 5$. The optimal choice $\lambda_R = 1.375$ shows excellent convergence properties through sixth order, while the trends to a consistent value across all $\lambda_R$ validate the renormalized perturbation expansion.
        The dashed black line shows the estimate at $\lambda_R$ = 1.375, and the gray shaded region indicates the final result with uncertainty.}
    \label{fig:meff_lambda_conv}
\end{figure}

Figure~\ref{fig:meff_lambda_conv} demonstrates the behavior of the effective mass ratio $m^*/m$ with perturbation order $N$ for various values of the screening parameter $\lambda_R$ at $T/T_F = 1/40$. The screening parameter $\lambda_R$ serves as a variational parameter that improves the convergence properties of the renormalized perturbative series without altering the physical results when the series is summed to all orders.

Assessing the systematic uncertainty from series truncation requires careful consideration, as relying on the final two perturbation orders can be misleading for certain choices of the screening parameter, such as $\lambda_R=1.375$, where the series might exhibit fortuitously small corrections.
We find that a three-point analysis, based on the variation across the last three perturbation orders ($N=4,5,6$), provides a consistent estimate for the systematic uncertainty when applied across different, well-behaved screening parameters. Given the observed robustness of this empirical approach, we adopt it to determine our final systematic error, applying the analysis to the data for $\lambda_R$ values near our optimal choice of 1.375. This method ensures the reported uncertainty is grounded in the consistent behavior of the series rather than being biased by the potentially anomalous convergence of a single order.

Our analysis yields $m^*/m=0.979(2)$ at the working temperature $T/T_F=1/40$. This uncertainty estimate incorporates both statistical errors from Monte Carlo sampling in the sixth-order Feynman integral calculations and systematic uncertainties from the analysis described above. The analysis in Fig.~\ref{fig:meff_lambda_conv} demonstrates that our renormalized perturbation expansion provides a controlled framework for high-order calculations in the UEG.

\subsection{Finite-temperature analysis}
\label{appendix:B2}

\begin{figure}
    \centering
    \includegraphics[width=0.85\linewidth]{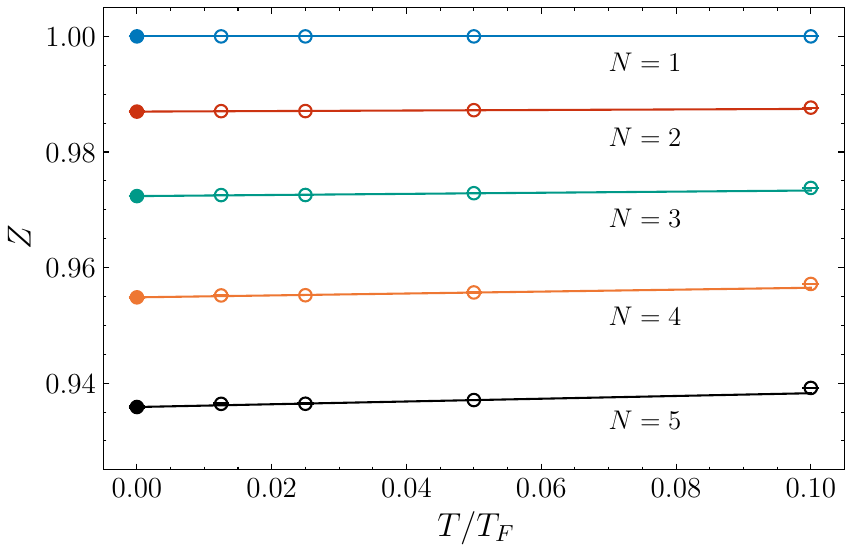}
    \includegraphics[width=0.85\linewidth]{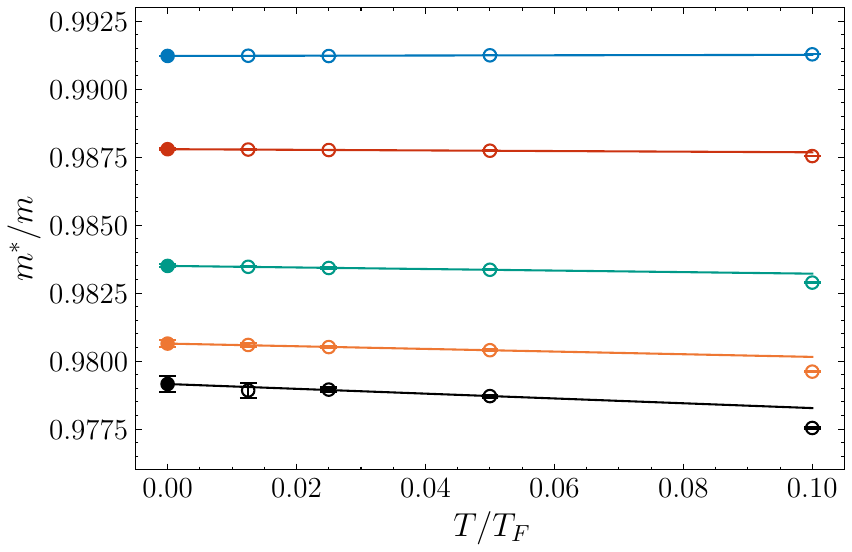}
    \caption{Temperature dependence of the quasiparticle weight $Z$ (upper panel) and effective mass ratio $m^*/m$ (lower panel) for $r_s = 5$ and $\lambda_R=1.375$ across perturbation orders $N = 1$--5. The observed linear-$T$ dependence enables extrapolation to zero temperature, while small thermal corrections at $T/T_F = 1/40$ confirm the accuracy of our working temperature choice.
    }
    \label{fig:finiteT_results}
\end{figure}

Figure~\ref{fig:finiteT_results} presents the temperature dependence of the quasiparticle weight $Z$ (upper panel) and effective mass ratio $m^*/m$ (lower panel) across perturbation orders $N = 1$--5 for the choice $\lambda_R=1.375$. Both quantities exhibit clear linear temperature dependence in the low-temperature regime $T< 0.1T_F $, consistent with the predictions of Landau Fermi liquid theory.

The linear temperature dependence arises from the analytical structure of the self-energy in Fermi liquids. Near the Fermi surface, the self-energy can be expanded as:
\begin{equation}
    \Sigma(k,\omega) = a \frac{k_F}{m} (k-k_{\rm F}) + b\omega+ ic (\omega^2+\pi^2T^2)+ \ldots \,,
\end{equation}
where $a$, $b$, and $c$ are real constants determined by the interaction strength.
The key insight is that the imaginary part $\mathrm{Im}\Sigma$ reflects phase-space constraints on particle-hole excitations.
At finite temperatures,  thermal broadening of the Fermi surface modifies these constraints, effectively replacing $\omega^2$ with $\max (\omega^2,\pi^2T^2)$ in this expansion. This temperature dependence in $\mathrm{Im}\Sigma$ propagates to the real part $\mathrm{Re}\Sigma$ through the Kramers-Kronig relations, introducing temperature-dependent corrections to both the frequency and momentum derivatives of $\mathrm{Re}\Sigma$.

The quasiparticle residue $Z$ is obtained from the frequency derivative of $\mathrm{Re}\Sigma$ as
\begin{equation}
    Z^{-1} = 1 - \left.\frac{\partial \mathrm{Re}\Sigma}{\partial\omega}\right|_{{\bf k}={\bf k}_\text{F},\,\omega=\mu} = 1-b \,.
\end{equation}
At finite temperatures, the parameter $b$ acquires a correction proportional to $T/T_F$ due to the thermal smearing of the Fermi surface, leading to
\begin{equation}
    Z(T) = Z(0)\left(1- \alpha_Z \frac{T}{T_F} + \ldots \right) \,,
\end{equation}
where $\alpha_Z$ is a dimensionless constant determined by forward scattering amplitudes.

The effective mass depends on both $Z$ and the momentum derivative of $\mathrm{Re}\Sigma$:
\begin{equation}
    \frac{m^*}{m} = Z^{-1} \left( 1+\frac{m}{k_F} \left.\frac{\partial \mathrm{Re}\Sigma}{\partial k}\right|_{{\bf k}={\bf k}_\text{F},\,\omega=\mu} \right) = \frac{1-b}{1+a} \,.
\end{equation}
The momentum derivative $\partial \mathrm{Re}\Sigma/\partial k$ samples the same thermal phase space as the frequency derivative, resulting in $a(T) -a(0) \sim T/T_F$. Combining the temperature dependence of both $Z$ and the momentum derivative yields:
\begin{equation}
    \frac{m^*(T)}{m} = \frac{m^*(0)}{m} \left(
    1+ \alpha_m \frac{T}{T_F} + \ldots  \right) \,,
\end{equation}
where $\alpha_m$ is an interaction-dependent constant that combines the contributions from both $b$ and $a$.
These linear-$T$ corrections represent universal features of Fermi liquid theory in the low-$T$ regime $T \ll T_F$, where quasiparticles remain well-defined excitations.

The linear-$T$ dependence observed across all perturbation orders in Fig.~\ref{fig:finiteT_results} validates the theoretical framework and enables extrapolation to $T = 0$. The small slopes demonstrate that thermal corrections remain modest throughout the studied parameter range for $\lambda_R=1.375$. The thermal correction to $m^*/m$ is approximately an order of magnitude smaller than our uncertainty of 0.002 at $T/T_F = 1/40$. This validates our computational strategy of performing high-order calculations at finite temperature rather than direct zero-$T$ computation, allowing our low-$T$ results to be directly compared with ground-state properties.

\bibliography{computationalgraph}

\end{document}